\documentclass[a4paper,11pt]{article}
\usepackage[left=2.5cm,right=2.5cm,top=2cm,bottom=2.5cm]{geometry}
\usepackage{epsfig,amsfonts,amssymb}
\usepackage{xcolor}
\usepackage{amsmath}
\usepackage{framed}
\usepackage{cite}
\usepackage{varioref}
\colorlet{shadecolor}{lightgray}
\usepackage{hyperref}
\usepackage{mathrsfs}
\usepackage{mathtools}
\usepackage{parskip}
\usepackage{soul}
\usepackage{enumitem}

\usepackage{placeins}  
\usepackage{flafter}   

\usepackage[utf8]{inputenc}
\usepackage{comment}
\usepackage{float, xcolor}
\usepackage{physics}
\usepackage{ragged2e}
\usepackage{booktabs}
\usepackage{latexsym}
\usepackage{mathtools}

\usepackage{subcaption}
\usepackage{graphicx}
\usepackage{cleveref}

\numberwithin{equation}{section}
\definecolor{darkgreen}{rgb}{0,0.5,0.25}
\definecolor{darkblue}{rgb}{0,0,0.6}
\usepackage{caption}
\usepackage{hyperref}
\hypersetup{
citecolor=red,
colorlinks=true,
filecolor=red,
linkcolor=blue,
linktocpage=true,
urlcolor=cyan
}

\def \be {\begin{equation}}
\def \ee {\end{equation}}
\def \bea {\begin{eqnarray}}
\def \eea {\end{eqnarray}}
\def \nn {\nonumber}
\newcommand{\eq}[1]{(\ref{#1})}

\begin{document}

\baselineskip 24pt

\begin{center}

{\Large \bf Post-Newtonian Effective Field Theory Approach to Entanglement Harvesting, Quantum Discord and Bell's Nonlocality Bound Near a Black Hole
}

\end{center}

\vskip .6cm
\medskip

\vspace*{4.0ex}

\baselineskip=18pt

\centerline{\large \rm Feng-Li Lin$^{a}$ and Sayid Mondal$^{b}$}

\vspace*{4.0ex}
\centerline{\  ~$^a$Department of Physics, National Taiwan Normal University, Taipei, 11677, Taiwan}

\centerline{  ~$^b$Instituto de Ciencias Exactas y Naturales,}
\centerline{ Universidad Arturo Prat, Playa Brava 3256, 1111346, Iquique, Chile}

\vspace*{1.0ex}
\centerline{\small E-mail: \href{fengli.lin@gmail.com}{fengli.lin@gmail.com} , \href{sayid.mondal@gmail.com}{sayid.mondal@gmail.com}}

\vspace*{5.0ex}

\centerline{\bf Abstract} \bigskip

Black holes, as characterized by the Hawking effect and Bekenstein-Hawking entropy, can be treated as a compact object carrying nontrivial quantum information obscured behind the event horizon. This quantum information, while hidden behind the event horizon, can be indirectly probed through the black hole's interactions with surrounding quantum fields. In this paper, we investigate how the quantum nature of a black hole influences the correlations harvested by a pair of static Unruh-DeWitt (UDW) detectors. To provide a comprehensive analysis, we employ various correlation measures: concurrence, quantum discord, and the violation of Bell's inequality, thereby shedding light on the quantum nature of the black hole as perceived by local probes. By treating the black hole as a tidally deformable thermal body under the quantum fluctuation of the mediator fields, as observed in \cite{Goldberger:2019sya, goldberger2020virtual, biggs2024comparing}, we employ a post-Newtonian effective field theory (PN-EFT) to derive the final states of the two UDW probes analytically. A key advantage of our approach is the ability to analytically derive all these correlation measures without encountering the complicated Matsubara sum of infinite thermal poles, as in the conventional approach based on quantum fields in curved spacetime. By tuning the relative strengths in the action of PN-EFT, we can extract the effects of the black hole on the entanglement harvesting, quantum correlation, and nonlocality bound of the UDW probe systems. Furthermore, our PN-EFT approach can be extended in future studies to include backreaction on black holes by accounting for higher-order PN corrections.

\vfill \eject

\baselineskip 18pt

\tableofcontents

\section{Introduction}

Black holes are simple yet nontrivial classical solutions to Einstein's theory of gravity. However, when combined with the quantum effects of quantum field theory in curved spacetime, they led to important discoveries such as Hawking radiation \cite{Hawking:1975vcx} and the Bekenstein-Hawking entropy \cite{Bekenstein:1973ur, Bekenstein:1974ax}. These new ingredients transform a classical black hole into a quantum one that resembles a thermal body and contains nontrivial quantum information, thereby resolving the information paradox \cite{PhysRevD.14.2460,Hawking1975}. Long-term efforts have been made to resolve the information paradox by studying the quantum nature of black holes using various approaches (see \cite{Page:1993wv,RevModPhys.93.035002,Mathur:2009hf,Penington:2019kki,Almheiri:2019qdq} and references therein). 

The information paradox suggests that quantum black holes may exist in non-trivial quantum states characterized by non-trivial quantum information measures, such as quantum entanglement. In the absence of a complete quantum theory of gravity to directly probe these mysterious quantum states of black holes, one may rely on probe systems that interact with black holes via mediator quantum fields to observe how their existence induces changes in their quantum information content. In this way, one may hope to peek at the mysterious quantum states of black holes. To extract additional intrinsic properties of black holes, one typically uses simple probes, such as the Unruh-DeWitt (UDW) detectors \cite{PhysRevD.14.870, DeWitt:1980hx}, which are point-like qubit systems. Employing these probes, one can study how the black hole destroys the quantum coherence of a superposition state of UDW probes in a which-paths experiment \cite{Danielson:2021egj, Danielson:2022sga, Danielson:2022tdw, Gralla:2023oya, biggs2024comparing, Wilson-Gerow:2024ljx}, or induces the entanglement among a pair of UDW probes \cite{Salton:2014jaa, Martin-Martinez:2015eoa, PhysRevD.93.044001, Henderson:2017yuv, Kukita:2017etu, Perche:2022ykt, gallock2021harvesting, Mendez-Avalos:2022obb, Lin:2024roh}. Along similar lines, one can also study the changes in other quantum information measures of the UDW probes to explore different aspects of the black hole's quantum nature.

In the conventional approach of studying quantum fields in curved spacetime, a black hole is usually treated as a background spacetime. The interactions between the black holes and the UDW probes are mediated by the surrounding quantum fields. To study the change in the information content of the UDW probes due to the presence of a black hole horizon, we need to obtain the probes' reduced states, from which we can calculate some quantum information measures. In this approach, the reduced states are dictated by the Wightman functions of the mediator quantum fields in the Hartle-Hawking vacuum. This conventional approach, which has been typically adopted in most studies, has two major disadvantages.  The first point is that black holes are compact objects; treating them as the spacetime background will miss situations in which they are moving, driven by external forces or backreaction. In fact, in the usual post-Newtonian (PN) approach \cite{Goldberger:2004jt, Goldberger:2007hy, Porto:2016pyg} by integrating out the weak gravity field, black holes are treated as point particles, with the gravitational effect encoded in higher-order interactions beyond the Newtonian approximation. The second disadvantage is that the Wightman functions of the mediator fields in the Hartle-Hawking vacuum are usually plagued by the Mastubara sum of an infinite number of thermal poles; this makes it hard to find the analytical forms for the reduced states of the UDW probes, which need delicate treatment even when just evaluating the reduced states numerically.  This obstacle also makes it difficult to compute other quantum information measures, apart from concurrence, which characterizes entanglement harvesting.

On the other hand, the universal feature of black holes shown in the Hawking effect and Bekenstein-Hawking entropy implies that the reduced states of UDW probes should also be universal, at least in the low-energy regime, i.e., when the black hole size is much smaller than the distance between probes and the black hole, and also the interaction time scale. Motivated by this, we may begin with some effective field theory (EFT) by integrating out the mediator fields, treating the black holes as point-like objects on the same footing as the UDW probes. The only thing missing from this EFT approach, as described, is how to incorporate the black hole's internal degrees of freedom, which can explain the Hawking effect and the Bekenstein-Hawking entropy. Fortunately, these degrees of freedom have been introduced in \cite{Goldberger:2019sya, goldberger2020virtual} by requiring their Wightman functions to yield the Hawking effect dictated by the Wightman functions of the mediator fields. Furthermore, it was realized in \cite{biggs2024comparing} that these degrees of freedom can be understood as the multipole moments caused by the dynamical tidal deformations of black holes as a thermal body at Hawking temperatures under the action of the fluctuations of the surrounding mediator fields. Note that even black holes are known to be perfectly rigid under static external fields \cite{Damour:2009vw, Binnington:2009bb, kol2012black, Charalambous:2021mea, Hui:2021vcv, Combaluzier-Szteinsznaider:2024sgb, Kehagias:2024rtz}, they can be tidally deformed universally under low-energy dynamical external fields \cite{Charalambous:2021mea, ivanov2022black, ivanov2023vanishing, perry2023dynamical}. This low-energy universality also yields the universal reduced states of the UDW probes.

\begin{figure}[ht]
	\centering
	\begin{subfigure}{.4\textwidth}
		\centering
		\includegraphics[width=.9\linewidth]{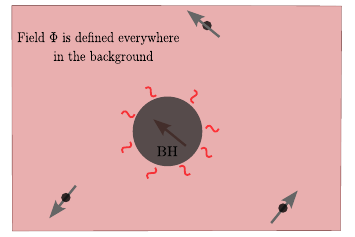}
		\caption{}
	\end{subfigure}\hspace{.45cm}
	\begin{subfigure}{.4\textwidth}
		\centering
		\includegraphics[width=.9\linewidth]{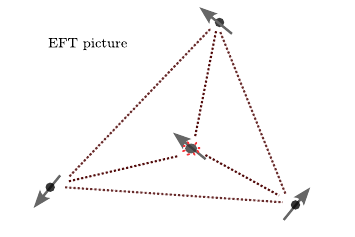}
		\caption{}
	\end{subfigure}
	\caption{ Comparison of the conventional approach (Left) based on quantum fields in curved spacetime and the PN-EFT approach (Right) in studying reduced states of UDW probes near a black hole. In the conventional approach, the black hole is treated as a background spacetime, which thermalizes the mediator field $\Phi$. However, in the PN-EFT approach, the black hole is treated as a point-like object with induced multipole moments and interacts with the point-like UDW probes via PN-type interactions, the higher-order terms of which characterize the back-reaction effects.}
\label{fig:EFT_pic}
\end{figure}

In this paper, we employ the PN-EFT approach to investigate changes in the quantum information content of two static UDW probes arising from their interactions with the black hole's induced degrees of freedom. In \cref{fig:EFT_pic}, we schematically compare the conventional approach and the PN-EFT approach adopted for our study in this paper. We show that the low-energy universality of the Wightman function for the black hole's tidally induced dynamical multipole moments yields an analytical form for the reduced states of the UDW probe. This also enables a straightforward derivation of entanglement harvesting, eliminating the need to perform the Matsubara sum numerically, as in the conventional approach. Our results on the patterns of entanglement harvesting are consistent with those of previous studies \cite{Henderson:2017yuv}. This verifies the validity of the PN-EFT approach. Based on the simple form of the reduced states, we further study the change of quantum discord \cite{ollivier2001introducing, Henderson:2001wrr} and the nonlocality bound \cite{HORODECKI1995340,  Brunner:2013est, buhrman2010nonlocality} of the UDW probes to reveal the effects due to the presence of a black hole. Moreover, the PN-EFT action encompasses probe-probe interactions and probe-black hole interactions, with the strength controlled by the size of the multipole moments and the inter-distances between them. By appropriately tuning these parameters, we can observe how each interaction, or combinations of interactions, affects the entanglement, quantum correlations, and nonlocality bounds of the UDW probes. This helps to extract the effects of black holes on the quantum information content of probes.

\FloatBarrier

For comparison with the quantities obtained by the conventional approach based on quantum fields in curved spacetime, we will use the PN-EFT to compute entanglement harvesting \cite{PhysRevD.93.044001, Henderson:2017yuv} by the UDW probes from the environmental field. One advantage of the PN-EFT approach over the conventional approach is that it yields nearly all results analytically. Furthermore, we can observe how the different interaction terms in the PN-EFT influence the pattern of entanglement harvesting. For example, suppose that we neglect the direct interaction between the multipole moments of two UDW probes. In that case, the remaining direct interactions between the UDW probes and the black hole will always yield no entanglement harvesting. This implies that direct interaction between UDW is relevant to reproducing the effect arising from their interaction with the mediator fields and can be seen as an interesting result of PN-EFT. 

In addition to entanglement harvesting, we consider quantum discord and the nonlocality bound in Bell-type experiments to characterize the impact of each interaction term in the PN-EFT on these quantum information measures. The quantum discord \cite{ollivier2001introducing, Henderson:2001wrr} characterizes non-classical (or pure quantum) correlations of bipartite states. The nonlocality bound \cite{HORODECKI1995340,  Brunner:2013est, buhrman2010nonlocality}, on the other hand, examines whether any PN-EFT interaction, such as the direct Coulombic interactions, can yield nonlocality in the sense of the Bell inequality or not.  As elaborated, in the PN-EFT framework, one can study the effect of each interaction on various quantum information measures, particularly for interactions with the black hole.  These effects are information-type responses that probe black-hole quantum states via UDW detectors. Examining these effects may help to understand the black hole's quantum states.

The remainder of the paper is organized as follows. In \cref{sec2}, we describe the PN-EFT for the black hole and the static UDW probes by treating the black hole as a thermal body with the universal low-energy dissipative transport phenomenon \cite{biggs2024comparing}. In \cref{sec3}, we analytically derive the low-energy reduced final states of UDW probes for three setups: (I) without the black hole; (II) neglecting direct interactions between UDW probes; and (III) including all the leading-order PN interactions. Based on these reduced states for the three setups, we derive the corresponding entanglement harvestings in \cref{sec4}, the quantum discords in \cref{sec5}, and the nonlocality bounds in \cref{sec6}. Finally, we summarize our results in \cref{sec7} and provide a brief discussion.

\section{EFT for UDW probes and Blackbody}\label{sec2}

This section reviews and extends the EFT approach of \cite{biggs2024comparing} to incorporate the interaction between the two UDW probes and a blackbody via an ambient mediator field. Such a blackbody can be a quantum black hole or a generic blackbody at a fixed temperature. Based on this formulation, we aim to study entanglement harvesting, quantum correlations, and Bell's nonlocality bound between the UDW probes.

We consider two UDW probes and the blackbody, each endowed with multipole moments denoted by $Q^I_i$, where $I$ is the tensor index characterizing the types of multipole moments and $i=1, 2, B$ labels the two detectors and the blackbody, respectively. These multipole moments $Q^I_i$ act as sources for and interact via the corresponding ambient massless tensor fields  $\Phi_I$. Moreover, the blackbody is treated as a quantum object; hence, its multipole moments $Q^I_B$ are promoted to quantum operators, whose spectral properties govern the dynamical evolution. In contrast, we treat the multipole moments  $Q^I_{1,2}$ of the UDW probes as classical quantities for simplicity. In addition to the multipole moment, the UDW probes can also possess internal quantum spin $J_{a=1,2}$ acting on the associated spin Hilbert space ${\cal H}^S_{a=1,2}$. For example, we can think of the UDW detector as a realistic atom that carries both the quantum (iso)spin and the multipole moment and interacts with the Maxwell field. Similarly, the blackbody can also own quantum spin denoted by $J_B$, formed from the coherent fermion condensation. A black hole with quantum spin is not a Kerr black hole with classical angular momentum due to the self-rotation. 

At low energies, the dynamics of the above system can be described by the following EFT action
\be\label{S_EFT}
S=S_{\rm Blackbody} + \frac{1}{g_I^2} S_{\Phi} + S_{\rm UDW}^{\Omega}+ \sum_I \int dt \;    \Phi_I(t)   \sum_{i=1,2, B} Q^I_i(t)  \otimes J_i(t) \;.
\ee
Here, $S_{\rm Blackbody}$ and $S_{\rm UDW}$ are the kinetic actions of the point-like blackbody and UDW probes with energy gap $\Omega$, respectively. Since we focus exclusively on static cases in this work, these terms will not play a dynamic role in the analysis. The term $\frac{1}{g_I^2} S_{\Phi}$ is the kinetic action for the ambient mediator field $\Phi_I$. We follow the convention adopted in \cite{biggs2024comparing} by normalizing it by the square of the dimensionless universal coupling constant $g_I$. In this paper, we consider only massless fields $\Phi_I$, such as: the scalar field with coupling constant $g_0$, the Maxwell field with $g_1$ corresponding to the fine structure constant, and Einstein's gravity field with $g_2:=\frac{G_N}{R^2}$ where $G_N$ is Newton's constant and $R$ is the typical size of the blackbody. The last term of \eq{S_EFT} encodes the interactions between the field $\Phi_I$, the multipole moment $Q_i^I$, and the auxiliary internal quantum spin $J_i$. This interaction is crucial in extracting quantum information measures from the vacuum states of $\Phi_I$ by the quantum spins of the UDW probes.

We can integrate out the ambient fields to induce direct interaction among the multipole-moment operators of the same type. Moreover, due to the massless nature of $\Phi_I$, after subtracting the self-energy part, the leading order action of the post-Newtonian effective field theory (PN-EFT) reads \cite{Porto:2016pyg}
\be\label{S_EFT_f}
S_{\rm PN-EFT} = \sum_I g_I \int dt \; \Big[ \sum_{a=1,2} O^I_{aB}(t) J_a(t) \otimes J_B(t)  + O^I_{12}(t) J_1(t) \otimes J_2(t) \Big]\;.
\ee
The key ingredient of \eq{S_EFT_f} is dictated by the effective coupling operators  $O^I_{ij}(t)$'s, which are given by 
\be
O^I_{ij}(t) := \frac{Q_i^{I_1}N_{I_1 I_2}(\hat{r}_{ij}) Q^{I_2}_j}{r_{ij}^{2I+1}}\;, \qquad i,j=1,2,B\;,
\ee
where $I_1$ and $I_2$ are two different sets of the type $I$ indices, and the distance vectors  are
\be
\vec{r}_{ij}:=r_{ij} \hat{r}_{ij}=\vec{r}_i - \vec{r}_j\;, \qquad {\rm with} \qquad |\hat{r}_{ij}|=1\;.
\ee
The multipole-moment structure tensor $N^{I,I'}$ for the monopole, dipole, and quadrupole is given respectively by \cite{biggs2024comparing}
\bea
&& N(\hat{r})=1 \;, 
\\
&& N_{i,j}(\hat{r})= \delta_{ij}-3\hat{r}_i \hat{r}_j\;, 
\\
&& N_{ij,kl}(\hat{r})=2 \delta_{ik}\delta_{jl}+35 \hat{r}_i  \hat{r}_j \hat{r}_k  \hat{r}_l -20 \hat{r}_i \delta_{jk}  \hat{r}_l\;.
\eea
We observe that the quantum spins of different entities now interact directly with each other via Coulombic power-law interactions, which reflect the massless nature of the mediator fields.

For simplicity, we only consider the interactions of the same tensor types by assuming the correlators among different tensor types are negligible. Moreover, for the EFT of \eq{S_EFT} to be valid, we shall assume 
\be\label{low_energy_c}
\bar{r}_i \ll r_{a B} \ll T\;, \qquad a=1,2\;, \quad i=1,2,B\;,
\ee
where $\bar{r}_i$ is the typical size of the blackbody or UDW probes, and $T$ is the duration of interaction between UDW probes and the blackbody.

The action \eq{S_EFT_f} is the leading-order (0-PN) action, i.e., instantaneous and Coulombic, because we integrate out the off-shell mediator fields and hold the entities static without following their geodesics. According to the PN formalism \cite{Goldberger:2004jt, Goldberger:2007hy, Porto:2016pyg}, one can systematically derive higher-order PN interactions within the velocity expansion for moving entities. For example, the 1PN effective action in gravity is the famous Einstein-Infeld-Hoffman action \cite{EIH}. With such EFT, which can also include radiation reactions at half-integral PN orders, one can study various quantum information measures for moving UDW probes and black holes in their full dynamic aspect. As a preliminary study, we restrict our attention to the leading order-EFT action of \eq{S_EFT_f}. Before we can adopt this PN-EFT to study the reduced states of UDW probes, we have one more issue to resolve regarding the multipole moments associated with the black hole.

In the above description of PN-EFT, the multipole moment $Q_B^I$ of the blackbody can be either intrinsic or extrinsically induced by the source fields $\Phi_I$. However, if the blackbody is a black hole, it doesn't seem easy to understand such multipole moments as the intrinsic quantities due to the no-hair theorem. It raises the question of whether multipole moments can be extrinsically induced by tidal deformation in the ambient fields. The answer appears to be no, as it is known that black holes cannot be tidally deformed. However, this holds only for static ambient fields, not for dynamic ones. Thus, $Q_B^I$ can be understood as the multipole moments induced by the dynamical $\Phi_I$, so that they are related by the linear response relation characterized by the dynamical tidal Love number (TLN).

From now on, we will consider only the extrinsic $Q^I_B$ for either a blackbody or a black hole. In such a case, the linear response relation between $Q_B^I$ and $\Phi_I$ in the frequency domain is just the Kubo formula \cite{Kubo:1957mj, Kubo1957StatisticalMechanicalTO} for the tidal deformation:
\be
\langle Q_B^{I_1}(\omega) \rangle_{\Phi} = \chi^{I_2 I_2}(\omega) \Phi_{I_2}(\omega)\;,
\ee
where the response function  $\chi^{I_1 I_2}(\omega)$ is the Fourier transform of the retarded Green function\footnote{Note that our convention for the retarded Green function differs from the one used in \cite{biggs2024comparing} by an overall minus sign, so that the fluctuation-dissipation relation \eq{FDT} also differs by a sign accordingly, similarly for the minus sign in front of $B^{I_1I_2}$ of \eq{AB_coeff} to take care of the convention difference.} 
\be
G^{I_1 I_2}_R(t)=-i \Theta(t)\langle  [Q^{I_1}_B(t),Q^{I_2}_B(0)] \rangle \;.
\ee
If the blackbody is in a thermal state $\rho_B$ of inverse temperature $\beta$, e.g., the Hartle-Hawking state when considering a quantum black hole, then the imaginary part of $\chi^{I_1 I_2}(\omega)$ is the dissipation kernel due to thermal or quantum transport. By the Kubo-Martin-Schwinger (KMS) condition \cite{Kubo:1957mj, PhysRev.115.1342} on the thermal state, it leads to the fluctuation-dissipation theorem: 
\be\label{FDT}
S^{I_1 I_2}(\omega) = - 2 (n_b(\omega)+1) {\rm Im} \chi^{I_1 I_2}(\omega)\;, \qquad n_b(\omega)=\frac{1}{e^{\beta \omega} -1}\;,
\ee
where the fluctuation kernel $S^{I_1 I_2}(\omega)$ is the spectral density of $Q_B^I$ defined by 
\be\label{spectral_W}
S^{I_1 I_2}(\omega):=\int_{-\infty}^{\infty} dt e^{i \omega t} \langle Q^{I_1}_B(t) Q^{I_2}_B(0) \rangle\;,
\ee
with the Wightman function
\be
\langle Q^{I_1}_B(t) Q^{I_2}_B(0) \rangle := {\rm Tr}_B\big[ \rho_B Q^{I_1}_B(t) Q^{I_2}_B (0)\big]\;.
\ee

At low energies, $\chi(\omega)$ manifests a universal form (see the extensive checks in \cite{biggs2024comparing}) as follows
\be\label{AB_coeff}
\chi^{I_1 I_2}(\omega) =\Big[ A^{I_1 I_2} + {\cal O}\big((\beta \omega)^2\big) \Big] - i \Big[ B^{I_1 I_2} \beta \omega +{\cal O}\big((\beta \omega)^3\big) \Big]\;,
\ee
with $A^{I_1 I_2}$ real and $B^{I_1 I_2}>0$, and their detailed expression depends on the blackbody and the ambient fields. The positivity $B^{I_1 I_2}$ is a signature of the dissipative transport. From \eq{FDT}, this then implies that
\be\label{S0_f}
 S^{I_1 I_2}(\omega) = B^{I_1 I_2} (2 + \beta \omega ) + {\cal O}\big((\beta \omega)^2\big) \;.
\ee

In the case of the black hole, $A^{I_1 I_2}$ and $B^{I_1 I_2}$ are nothing but, respectively, the static and dynamical TLNs, which can be determined by the linearized field equation of $\Phi_I$ around the black hole spacetime \cite{Charalambous:2021mea, ivanov2022black, ivanov2023vanishing, perry2023dynamical}. By imposing the ingoing condition at the horizon for the radial solution $\Phi_I(r)$, one can determine the ratio between the coefficients of $1/r^{I+1}$ and $r^I$ terms at $r\rightarrow \infty$ to yield the TLNs. The result gives  $A^{I_1 I_2}=0$ and $B^{I_1 I_2}$'s take the form for a black hole with radius $r_B$
\begin{equation}\label{B_scaling}
B=c_0 \frac{r_B}{g_0}, \qquad B^{i j}=c_1 \frac{r_B^3}{g_1} \delta^{i j}, \qquad B^{i k j l}=c_2 \frac{r_B^5}{g_2}\left(\delta^{i k} \delta^{j l}+\delta^{i l} \delta^{j k}-\frac{2}{3} \delta^{i j} \delta^{k l}\right),
\end{equation}
where $c_{0,1,2}$ are some ${\cal O}(1)$ constants determined by the black hole background\footnote{For example, for a Schwarzschild black hole with the inverse of Hawking temperature $\beta = 4\pi r_B$, $c_0=1$, $c_1=\frac{1}{6}$ and $c_2=\frac{1}{360\pi}$ \cite{biggs2024comparing, goldberger2020virtual, Fabbri:1975sa, Page:1976df, Charalambous:2021mea, ivanov2022black}. Note that these coefficients may slightly differ from the ones in the literature due to the different conventions adopted for the fundamental units.}. However, their values will not affect the qualitative results obtained in this paper. 

We shall point out an important property that $S^{I_1 I_2}(0)$ is non-vanishing if $B^{I_1 I_2} \ne 0$.  It implies a finite portion of zero modes, which can facilitate the transport of quantum information, e.g., quantum entanglement, and yield nontrivial reduced final states of UDW probes, as we will see later. From \eq{AB_coeff}, we notice that the dissipation kernel ${\rm Im} \chi^{I_1 I_2}(0)=0$. However, due to the appearance of $n_b(\omega)$ in \eq{FDT}, and the fact that $n_b(\omega) \sim \frac{1}{\beta \omega}$ as $\omega \rightarrow 0$, one then have $S^{I_1 I_2}(0)= 2 B^{I_1 I_2}$. In the case of the black hole, this combines the classical tidal deformation and the quantum effect of Hawking radiation.

\section{Reduced Dynamics of UDW probes}\label{sec3}

To study the reduced dynamics of the static UDW probes under the influence of the ambient field around a blackbody such as a black hole or a thermal body at a fixed temperature, we start with an initial state $\rho^i =\rho^i_{12}\otimes \rho_B$ with $\rho^i_{12}$ a pure product state in $\bigotimes_{a=1,2}{\cal H}^S_a$ for the quantum spins of UDW probes and $\rho_B$ the blackbody's thermal state. Without loss of generality, in this paper, we will always assume $\rho^i_{12}=|0\rangle_1\otimes 0\rangle_2 \langle0|_1 \otimes \langle0|_2=|00\rangle\langle 00| $ with $|0\rangle_a$ being the lowest weight state of spin $J_a$. The thermality of the black hole is due to the quantum effect of $\Phi_I$ in a curved background, i.e., Hawking radiation. In our PN-EFT, this thermal feature is inherited by $\rho_B$ and encoded in the spectral density $S^{I_1 I_2}(\omega)$.

The reduced final state of the UDW probes for their lifetimes along the whole worldlines can be obtained by
\be\label{rho12_1}
\rho_{12}^f={\rm Tr}_B \Big[U_{w_T} \big(\rho_{12}^i \otimes \rho_B \big) U^{\dagger}_{w_T} \Big]\;,
\ee
with the evolution operator in the interaction picture constructed from the EFT action \eq{S_EFT_f}:
\be\label{U_generic}
U_{w_T} ={\cal T}\;e^{ - i \sum_I g_I \int_{-\infty}^{\infty} dt \; w^I_T(t) \big[ \sum_{a=1,2} O^I_{aB}(t) J_a(t) \otimes J_B(t) + O^I_{12}(t) J_1(t)\otimes J_2(t) \big]} \;,
\ee
where $w_T^I(t)$ is a switching function with an effective time interval $T$, i.e., $\int_{-\infty}^{\infty} \vert w_T^I(t) \vert^2 dt =T$,  to characterize the turn-on profile of the type-$I$ interaction. Conventionally, one chooses the Gaussian-type switching function for some time interval during which the interaction is effectively turned on. We must note that $T$ is required to satisfy \eq{low_energy_c} for the EFT to be valid.

To proceed, we select a specific setup to simplify the calculation while preserving the key ingredients. We consider no quantum spin for the blackbody and choose the UDW probes as identical two-level qubits 
with the energy gap $\Omega$, that is
\be
J_a=e^{i\Omega t} \sigma_a^+ + e^{-i \Omega t} \sigma_a^-\;,
\ee
where $\sigma^+_a|0\rangle_a=|1\rangle_a$,  $\sigma^-_a|1\rangle_a=|0\rangle_a$ and $(\sigma_a^+)^2=(\sigma_a^-)^2=0$ for $a=1,2$.  

\subsection{Case I: Without black hole}

We begin by examining a seemingly simple scenario: the absence of a black hole, by setting $Q_B^I$ to zero. This is simply the case with two static UDW detectors interacting via quantized Coulombic mediator fields, which are sourced by the corresponding charge, dipole moment, or quadrupole moment, as discussed previously \footnote{In this paper, we consider the {\it quantum} mediator fields, even for the (linearized) gravity. One can find in \cite{Lin:2025ipk} the cases with classical mediator fields and their (in) ability to entangle the well-separated UDW detectors.}.


In this simplified setting, the time-evolution operator \eq{U_generic} of a specific type $I$ in this case is reduced to 
\be\label{U_EFT}
U_{w_T}={\cal T}\;e^{ - i g_I \int_{-\infty}^{\infty} dt \; w_T(t) O^I_{12}(t) J_1(t)\otimes J_2(t) \big]} \;.
\ee

We assume the two-detector system is initially in its ground state, $|\psi\rangle_i=|00\rangle$. Under the action of the operator \eqref{U_EFT}, the system evolves to the following final state \footnote{Expanding $J_1(t) \otimes J_2(t)$ we obtain $e^{2 i \Omega t} \sigma_1^{+} \sigma_2^{+}+e^{-2 i \Omega t} \sigma_1^{-} \sigma_2^{-}+\sigma_1^{+} \sigma_2^{-}+\sigma_1^{-} \sigma_2^{+}$. Acting on $|00\rangle$, the mixed terms $\sigma_1^{+} \sigma_2^{-}$and $\sigma_1^{-} \sigma_2^{+}$annihilate the state, and $\sigma_1^{-} \sigma_2^{-}$vanishes because $\sigma_a^{-}|0\rangle_a=0$. Thus the only surviving contribution is the pair-excitation term $e^{2 i \Omega t} \sigma_1^{+} \sigma_2^{+}|00\rangle=e^{2 i \Omega t}|11\rangle$, so the dynamics is confined to the $\{|00\rangle,|11\rangle\}$ subspace.}:
\be\label{UDW_fina_pure}
 |\psi\rangle_f=\cos \theta|00\rangle-i \sin \theta|11\rangle\;,
\ee
where the parameter $\theta$ is expressed as
\be\label{para_theta}
\theta=g_I O_{12}^II_{\omega}\;,\quad I_w=\int_{-\infty}^{\infty} w_T(t) e^{i 2 \Omega t} d t\;.
\ee

To ensure  the finite interaction time, we employ the Gaussian switching function:
\be\label{window_G}
w_T(t)=\frac{e^{-\frac{t^2}{4 T^2}}}{(2 \pi)^{1/4}}\;,
\ee
such that $\int_{-\infty}^{\infty} \vert w_T(t) \vert^2  dt = T$. With this choice, one obtains \begin{equation} 
I_w = (\sqrt{2}\pi)^{1/4}\, T\, e^{-4\Omega^2 T^2}\;.
\end{equation}


Even though the initial state is a product state, the final state is generally an entangled state with Schmidt rank two except for the {\it magic points} with $\theta = n \pi/2$ with $n\in \mathbb{Z}$.

The resulting density matrix for the final state $\rho_{12}^f=|\psi\rangle_f\langle\psi|$ takes the form
\be\label{red_final_noBH}
\rho_{12}^f =\left(
\begin{array}{cccc}
 \cos ^2\theta  & 0 & 0 & i \sin \theta  \cos \theta  \\
 0 & 0 & 0 & 0 \\
 0 & 0 & 0 & 0 \\
 -i \sin \theta  \cos \theta  & 0 & 0 & \sin ^2\theta  \\
\end{array}
\right)\;.
\ee

\subsection{Case II: Neglecting direct interaction between UDW probes}
In this case, we assume the condition
\be\label{no_backr}
|Q^I_{1,2}|\ll |Q^I_B|\;, \quad {\rm or} \quad |r_{12}|\gg |r_{1B}|, |r_{2B}|\;,
\ee
so that the term of $O^I_{12}(t) J_1(t)\otimes J_2(t)$ in \eq{U_generic} can be neglected compared to the $\sum_{a=1,2} O^I_{aB}(t) J_a(t)$. We are interested in the relevance of mutual interactions among UDW probes to reproducing entanglement-harvesting patterns using the conventional approach. The evolution operator of \eq{U_generic} given a specific type $I$ is  reduced to 
\be \label{U_simp}
U_{w_T} ={\cal T}\;e^{- i g_I \int_{-\infty}^{\infty} dt \; w_T(t)  \sum_{a=1,2} O^I_{aB}(t) J_a(t) }\;.
\ee
To obtain the final reduced state of UDW probes, we also assume no background value of $Q^I_B$, i.e., $\langle Q_B^I \rangle=0$. It is just the statement of the no-hair theorem if the blackbody is a black hole. Up to ${\cal O}\left({g_I \over r^{2(2I+1)}_{ij}}\right)$, the result takes the form of $X$-state \cite{yu2005evolution} as follows:
\begin{equation}\label{dtec_den_mat}
\rho_{12}^f =\left(\begin{array}{cccc}
1- P_1 -  P_2 & 0 & 0 &  X \\
0 & P_2 & C & 0 \\
0 &  C^* &  P_1 & 0 \\
 X^* & 0 & 0 & 0
\end{array}\right)+\mathcal{O}\left({ g^2_I \over r^{4(2I+1)}_{ij}} \right),
\end{equation}
where
\bea\label{P_a_i}
&& P_a := g^2_I {M_{a I_1} M_{a I_2}\over r^{2(2I+1)}_{a B} } P^{I_1 I_2}_T(\Omega)\;,
\\
&& C :=  g^2_I {M_{1 I_1} M_{2 I_2}\over r^{2I+1}_{1B} r^{2I+1}_{2B}}  P^{I_1 I_2}_T(\Omega) \;, \label{C_i}
\\
&& X := - g^2_I {M_{1 I_1} M_{2 I_2}\over r^{2I+1}_{1B} r^{2I+1}_{2B}} X^{I_1 I_2}_T(\Omega)\;, \label{X_i}
\eea
with 
\be
M_{a I_1}:=Q_a^{I}N_{I I_1}(\hat{r}_{aB})\;,
\ee
and
\bea\label{P_I1_I2_0}
P^{I_1 I_2}_T(\Omega):= \int_{-\infty}^{\infty} dt \int_{-\infty}^{\infty} dt' w_T(t) w_T(t') \; e^{-i\Omega (t-t')} \langle Q^{I_1}_B(t) Q^{I_2}_ B(t') \rangle\;,
\\ \label{X_X}
X^{I_1 I_2}_T(\Omega):=\int_{-\infty}^{\infty} dt \int_{-\infty}^{\infty} dt' w_T(t) w_T(t') \; e^{-i\Omega (t+t')} \langle Q^{I_1}_B(t) Q^{I_2}_ B(t') \rangle\;.
\eea 
Note that this form of $X$-state has also been obtained in \cite{Martin-Martinez:2015eoa, PhysRevD.93.044001, Henderson:2017yuv} by the conventional approach. Moreover, from \eq{P_a_i} and \eq{C_i} we have a exact relation
\be\label{C2=PP}
C^2=P_1 P_2\;,
\ee
which will be helpful when evaluating certain {quantum information} measures later on. 

Adopting the Gaussian window function \eq{window_G} for $w_T(t)$, and using the following fact when performing the change of time variables by $t_{\pm}=t\pm t'$
 \be
\int_{-\infty}^{\infty} dt \int_{-\infty}^{\infty} dt' \; w_T(t) w_T(t') \; e^{-i\Omega (t \pm t')}={1\over 2 } \int_{-\infty}^{\infty} dt_- \int_{-\infty}^{\infty} dt_+  \; w_T({t_- \over \sqrt{2}}) w_T({t_+ \over \sqrt{2}}) e^{-i \Omega t_{\pm}}\;,
\ee
we can simplify \eq{P_I1_I2_0} and \eq{X_X} into 
\bea
P^{I_1 I_2}_T(\Omega) = (2 \pi)^{1/4} T \int_{-\infty}^{\infty} dt_- \; e^{-i \Omega t_-} w_T({t_- \over \sqrt{2}}) \langle Q^{I_1}_B(t_-) Q^{I_2}_ B(0) \rangle\;,
\\
X^{I_1 I_2}_T(\Omega) = (2 \pi)^{1/4} T e^{-\Omega^2 T^2} \int_{-\infty}^{\infty} dt_- \;  w_T({t_- \over \sqrt{2}}) \langle Q^{I_1}_B(t_-) Q^{I_2}_ B(0) \rangle \;.
\eea
In the above, we have used the fact that $\langle Q^{I_1}_B(t) Q^{I_2}_ B(t') \rangle$ is time-translation invariant so that $\langle Q^{I_1}_B(t) Q^{I_2}_ B(t') \rangle = \langle Q^{I_1}_B(t_-) Q^{I_2}_ B(0) \rangle$.

Furthermore, using the convolution theorem, we have
\be\label{convolute}
\int_{-\infty}^{\infty} dt_- \; e^{-i \Omega t_-} w_T({t_- \over \sqrt{2}}) \langle Q^{I_1}_B(t_-) Q^{I_2}_ B(0) \rangle = \sqrt{2} \int_0^{\infty} d\omega \; \tilde{w}_T[\sqrt{2}(\omega - \Omega)] S^{I_1 I_2}(\omega)\;,
\ee
where the spectral density $S^{I_1 I_2}(\omega)$ is assumed to be bounded below, i.e., it vanishes for $\omega < 0$, and  the Fourier transform of $w_T(t)$ is
\be
\tilde{w}_T[\omega]=(8\pi)^{1/4} T e^{-\omega^2 T^2}\;. 
\ee
Thus, $\tilde{w}_T[\sqrt{2}(\omega - \Omega)]$ is roughly a window function in frequency domain centered around $\omega=\Omega$ with the window size of ${\cal O}(1/T)$.  We are in the low-energy regime constrained by \eq{low_energy_c} so that $S^{I_1 I_2}(\omega)$ for the black body is nearly $\omega$-independent as discussed in \eq{S0_f}, i.e., $ S^{I_1 I_2}(\omega) \simeq S^{I_1 I_2}(0)$ if $\beta \omega \ll 1$. Thus, combining all the above, $P_T^{I_1 I_2}$ and $X_T^{I_1 I_2}$ can be approximated to be
\bea
P_T^{I_1 I_2}(\Omega) &\simeq &  \pi T \big[ 1 + {\rm erf}(\sqrt{2} \Omega T) \big] S^{I_1 I_2}(0) \;,
\\
X_T^{I_1 I_2}(\Omega) & \simeq &  \pi T e^{-2 \Omega^2 T^2} S^{I_1 I_2}(0) \;.
\eea
It implies that 
\be\label{X<P}
X_T^{I_1 I_2}(\Omega) \le P_T^{I_1 I_2}(\Omega)\;,
\ee
Thus, also by \eq{P_a_i} and \eq{X_i} it yields
\be \label{X2lePP}
|X|^2 \le P_1 P_2\;.
\ee

We shall remark that the reduced final state of \eq{dtec_den_mat} is evaluated up to ${\cal O}(g_I)$ instead of ${\cal O}(g^2_I)$. Formally, $P_a$, $C$ and $X$ are ${\cal O}(g^2_I)$ as shown in \eq{P_a_i} to \eq{X_i}. However, they are in fact ${\cal O}(g_I)$ because $P_T^{I_1 I_2}$ and $X_T^{I_1 I_2}$ contain a factor of $1/g_I$ inherited from $B^{I_1 I_2}$ of \eq{B_scaling}, which is the low-energy expression of $S^{I_1 I_2}(\omega)$, i.e., \eq{S0_f}.

\subsection{Case III: Including all direct interactions}

Finally, in our third example, we consider the full evolution operator \eq{U_generic} of a particular $I$-type without any truncation as adopted in the previous two examples. The reduced final state takes the form of the full $X$-state as follows:
\begin{equation}\label{rho_full}
\rho_{12}^f =\left(\begin{array}{cccc}
1- P_1 -  P_2 - P_{12} & 0 & 0 &  X+X_0 \\
0 & P_2 & C & 0 \\
0 &  C^* &  P_1 & 0 \\
 X^*+X^*_0 & 0 & 0 & P_{12}
\end{array}\right)+\mathcal{O}\left({ g^3_I \over r^{4(2I+1)}_{ij}} \right),
\end{equation}
where
\be
P_{12}=X_0 X_0^* \;,
\ee
and the other variables are defined earlier.

We now consider the dependence of the matrix elements in \eq{rho_full} on the scales such as $T$, $\bar{r}_i$, $r_{ij}$ and $\Omega$. Note that these scales should satisfy \eq{low_energy_c} for the low-energy EFT to be valid. 
From dimensional analysis:
\be
\vert Q_a^I \vert = q_a \bar{r}_a^{I}\;, \qquad a=1,2\;,
\ee
where the dimensionless $q_a$ denotes the strength of the corresponding multipole moments. 
Along with the scaling relation of $B^{I_1 I_2}$ of \eq{B_scaling}, the scale dependence can be expressed as 
\bea
&& P_a \sim g_I q_a^2 \big({\bar{r}_a \over r_{aB}}\big)^{2I} \big({\bar{r}_B \over r_{aB}}\big)^{2I+1} {T\over r_{aB}}\;, \qquad 
 X_0 \sim g_I q_1 q_2  \big({\bar{r}_1 \bar{r}_2 \over r_{12}^2 }\big)^I {T\over r_{12}} e^{- 4 \Omega^2 T^2}\;, 
\\
&& C\sim  g_I q_1 q_2  \big({\bar{r}_1 \bar{r}_2 \over r_{1B} r_{2B}}\big)^I \big({\bar{r}^2_B \over r_{1B}r_{2B}}\big)^I \big({\bar{r}_B T\over r_{1B} r_{2B} } \big)\;,
\qquad  X \sim C e^{-2 \Omega^2 T^2} \;, \qquad P_{12}=\vert X_0 \vert^2 \;. \; \qquad 
\eea
Note that $P_{12}$ is one order higher in $g_I$ than the other elements. Thus, we can neglect it if we restrict to $g_I\ll 1$ by considering just the leading-order EFT. Based on the above scaling relation, we have
\be\label{X0_to_X}
{X_0 \over X} \sim  \big({r_{1B} r_{2B} \over r^2_{12} }\big)^I \big({r_{1B} r_{2B} \over \bar{r}^2_B }\big)^I  \big({r_{1B} r_{2B} \over r_{12} \bar{r}_B }\big)  e^{-2 \Omega^2 T^2} \sim  \big({r_{aB} \over \bar{r}_B }\big)^{2 I+1}e^{-2 \Omega^2 T^2} \;.
\ee
In arriving last expression, we have assumed $r_{1B}\simeq r_{2B} \simeq r_{12}$. To ensure \eq{U_simp} and \eq{dtec_den_mat}, we have assumed \eq{no_backr}. With the help of \eq{X0_to_X} and the conditions \eq{low_energy_c} for PN-EFT to be valid, we can now turn \eq{no_backr} into a more precise one by requiring $X_0 \ll X$. From \eq{low_energy_c}, one should require $r_{a B} \gg \bar{r}_B$, so that $X_0 \ll X$ holds only if $\bar{r}_B$ or $\Omega T$ is sufficiently large, i.e., 
\be
X_0 \ll X \quad \Longrightarrow \quad \bar{r}_B^{2I+1} \gg r_{aB}^{2I+1} e^{-2 \Omega^2 T^2}  \quad {\rm or} \quad 2 (\Omega T)^2 \gg (2I +1) \ln\big({r_{aB} \over \bar{r}_B }\big)\;. 
\ee
It can, however, be achieved without any problem in our setup.

\subsection{Features of reduced final states and {quantum information} measures}
The properties of the final state of the UDW detectors, denoted as $\rho_{12}^f$, differ significantly across the three cases considered.
\begin{enumerate}
    \item Case I: The final state is pure, as the evolution is unitary.
    \item Cases  II and  III: The final reduced state is a mixed state due to tracing out the black hole's Hilbert space. Specifically, it belongs to the class of $X$-states, which are characterized by the parameters $X, X_0, P_1$, and $P_2$. A key constraint for these cases is that the parameters are related by $|C|^2=P_1 P_2$ due to the positivity of the density matrix.
\end{enumerate}

An useful way to characterize these $X$-states is through their eigenvalues $\left\{\lambda_{0,1, \pm}\right\}$, which are accurate up to order $\mathcal{O}(g)$ :
\be\label{rho_f_eigen_1}
\lambda_{0,1, \pm}=0, \quad 1-P_1-P_2, \quad \frac{1}{2}\left(P_1+P_2 \pm \sqrt{\left(P_1-P_2\right)^2+4|C|^2}\right)\;.
\ee
These eigenvalues are notably independent of the parameters $X$ and $X_0$, which means the eigenvalue spectrum is identical for both Case II and Case III.

By applying the constraint $|C|^2=P_1 P_2$, the eigenvalues simplify significantly to:
\be
\lambda=0, \quad 1-P_1-P_2, \quad P_1+P_2, \quad 0\;.
\ee
The presence of two zero eigenvalues reveals that the final state $\rho_{12}^f$ for Cases II and III is, at most, a rank-two state.

In this work, to study the nature of correlations present in the final state, we will utilize the following quantum information measures:
\begin{enumerate}
    \item To quantify the entanglement harvesting from the quantum field, we compute concurrence. For cases II and III, the final reduced state is a mixed $X$-state, and we evaluate concurrence ${\cal C}$ to characterize the entanglement harvesting \cite{PhysRevLett.80.2245} from the ambient quantum field. For $X$-states considered in this paper, the concurrence is given by \cite{PhysRevD.93.044001, Henderson:2017yuv} 
    \be\label{concurrence}
    {\cal C}(\rho^f_{12})= 2 \; {\rm max}\big\{ 0, \vert X \vert - \sqrt{P_1 P_2} \big\}\;.
    \ee

    \item To characterize non-classical correlation, we compute Quantum discord \cite{ollivier2001introducing, Henderson:2001wrr}, denoted as ${\cal D}$. For the pure state (Case I), ${\cal D}$ coincides with the entanglement entropy. For the $X$-states, ${\cal D}$ is independent of the terms $(\rho_{12}^{f})_{14}=(\rho_{12}^{f})_{41}$, i.e., $X$ and $X_0$. Consequently, ${\cal D}$ is identical for Cases II and III up to ${\cal O}(g)$.  

    \item Bipartite nonlocality bound $S_{\rho}$ for CHSH-type experiment \cite{HORODECKI1995340,  Brunner:2013est, buhrman2010nonlocality} characterizes the nature of correlations. It depends on all the parameters in $\rho^f_{12}$ of $X$-states considered.  
    
\end{enumerate}

In the subsequent sections, we provide a detailed analysis of the aforementioned quantum information measures for the final state of the two-detector system across all three cases.  For simplicity, when presenting the numerical plots of the quantum information measures, we will consider only the scalar-type environmental field; that is, the tensor index $I$ will be absent. Also, we choose $g_0=0.01$ to ensure the validity of perturbative results, along with fixing $\bar{r}_B=\bar{r}_1=\bar{r}_2=1$, $q_1=q_2=1$ for convenience. We also fix a tiny energy gap $\Omega=0.001$, so that $T$ can be large enough to satisfy the condition \eq{low_energy_c} for the validity of PN-EFT. Thus, the quantum information measures shown below are just the functions of the interaction time $T$, the inter-distances $r_{12}$ and $r_{1B}$. Finally, to ensure the physical validity of our results, our analysis is restricted to parameter regimes in which the final reduced state $\rho_{12}^f$ is a valid density matrix. This is ensured by verifying that its purity remains within the physical bound, i.e., $\operatorname{Tr}\left(\rho_{12}^f\right)^2 \leq 1$.

\section{Entanglement Harvesting}\label{sec4}

Given an unentangled initial state of the two-detector system, the entanglement harvesting due to the mutual interactions among detectors and black holes can be measured by concurrence which is 
an entanglement monotone and equal zero if the bipartite states are separable (unentangled).

\subsection{Case I}
\begin{figure}[ht]
	\centering
	\begin{subfigure}{.4\textwidth}
		\centering
		\includegraphics[width=1\linewidth]{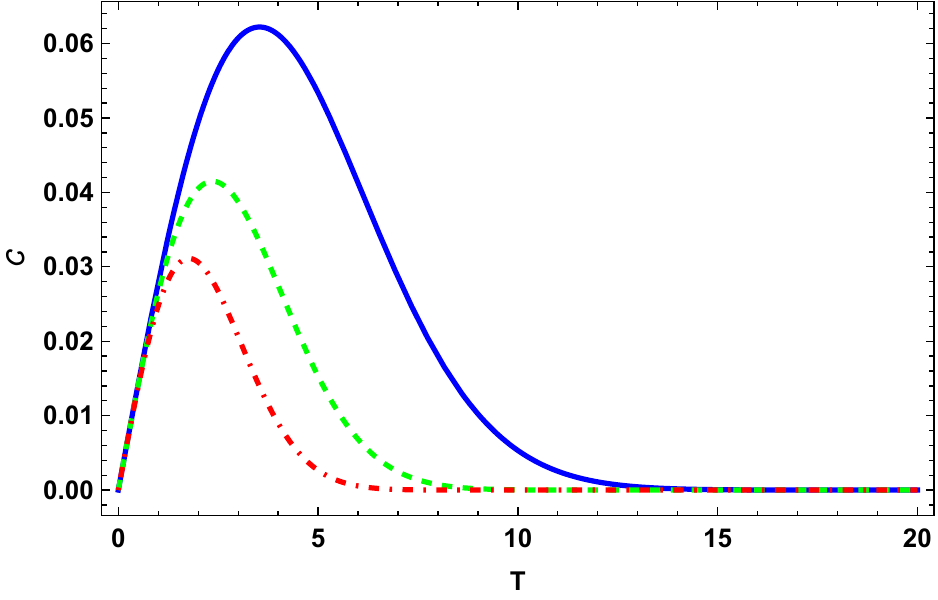}
		\caption{}
		\label{fig_C1_1}
	\end{subfigure}\hspace{.45cm}
	\begin{subfigure}{.4\textwidth}
		\centering
		\includegraphics[width=1\linewidth]{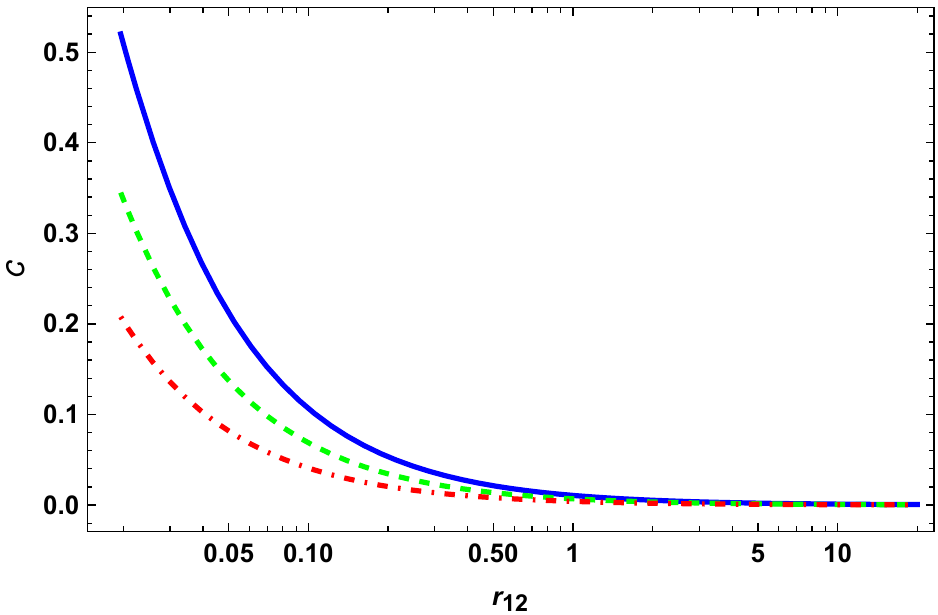}
		\caption{}
		\label{fig_C1_2}
	\end{subfigure}
	\caption{Concurrence $\mathcal{C}$ for $\rho^f_{12}$ of \eq{red_final_noBH}: (a) $\mathcal{C}$ vs $T$ for various $r_{12}$: $r_{12}=10$ (solid-blue), $r_{12}=20$ (green-dashed) and $r_{12}=30$ (red-dot-dashed). (b) $\mathcal{C}$ vs $r_{12}$ for various $T$: $T=800$ (solid-blue), $T=850$ (green-dashed) and $T=900$ (red-dot-dashed). In this figure and the other ones shown later on, we fix $g_0=0.01$, $\Omega=0.001$, and $\bar{r}_B=\bar{r}_1=\bar{r}_2=1$, $q_1=q_2=1$.}
\label{fig_con_BD_mono_vs_w}
\end{figure}

In this case, the final state of the two-detector system, $\rho^f_{12}$, is a pure state given by \eq{red_final_noBH} due to the absence of the black hole. We quantify the entanglement harvesting by measuring concurrence, which reads
\begin{equation}
\mathcal{C}=\left| \sin 2\theta \right| = \left|\sin \left(\frac{2 g_0 q_1 q_1 (\sqrt{2} \pi)^{1 / 4} T e^{-4 \Omega^2 T^2}}{r_{12}}\right)\right|\;,
\end{equation}
where we have used the definition of the parameter $\theta$ in \eq{para_theta}. A non-zero value of $\mathcal{C}$ indicates that the pair of UDW detectors can successfully harvest entanglement from the vacuum, a process mediated by their mutual Coulombic interaction. The behavior of the concurrence is depicted in \cref{fig_con_BD_mono_vs_w}. Notably, the amount of harvested entanglement reaches a maximum value when the interaction time and the energy gap approach $\Omega T \sim {\cal O}(1)$. This condition corresponds to an energy-conserving process, which enhances the probability of detector transitions and thus optimizes the transfer of quantum correlations from the field to the detectors. As noted before, there are magic points $\theta=n\pi/2$ with $n \in \mathbb{Z}$ where the oncurrence vanishes, and they deserve further attention. 

\subsection{Case II}

We now turn to the second case of entanglement harvesting, specifically in the presence of a black hole, while neglecting the direct interaction between the UDW probes. As mentioned before, due to the relation \eq{X2lePP}, i.e., $|X|^2 \le P_1 P_2$, in the low energy regime $\beta\omega \ll 1$, we can conclude
\be
{\cal C}(\rho^f_{12}) =0 \quad {\rm for} \quad \rho_{12}^f = \eq{dtec_den_mat}\;.\label{C2}
\ee
We can go to ${\cal O}(\beta \omega)$ of \eq{S0_f}, and the above result will not change. This result implies that one cannot harvest entanglement from a quantum blackbody, such as a quantum black hole, in the low-energy regime. Our findings contrast with those obtained via the conventional approach, which predicts non-vanishing entanglement harvesting from the ambient-field thermalized by the event horizon \cite{Henderson:2017yuv, Lin:2024roh, Koga:2019fqh}. This suggests that entanglement harvesting near the event horizon may arise from the high-energy regime. However, as we will demonstrate in the subsequent case, these features of entanglement harvesting can be recovered by incorporating the mutual Coulombic interaction between the UDW probes. This implies that all the EFT interactions at the same PN order are relevant toward reproducing the full result obtained from the conventional approach.

\subsection{Case III}
This subsection considers entanglement harvesting for the third case, in which all EFT interactions are included. The harvested entanglement is quantified by concurrence, which is given as
\be\label{concurrence3}
{\cal C}(\rho^f_{12})= 2 \; {\rm max}\big\{ 0, \vert X+X_0 \vert - \sqrt{P_1 P_2} \big\} \quad {\rm for} \quad \rho_{12}^f =\eq{rho_full} \;.
\ee

\FloatBarrier
\begin{figure}[ht]
	\centering
	\begin{subfigure}{.3\textwidth}
		\centering	\includegraphics[width=1.\linewidth]{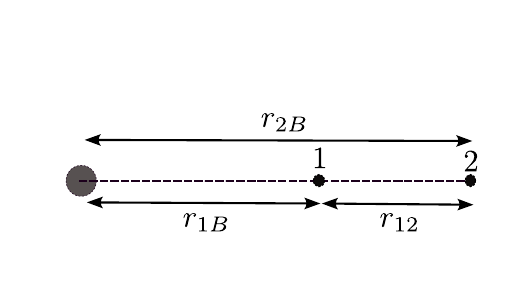}
		\caption{Line-up configuration with a black hole at the leftmost.}
		\label{fig:BH_UDW_C31}
  \end{subfigure}\hspace{.5cm}
	\begin{subfigure}{.3\textwidth}
		\centering	
\includegraphics[width=1.\linewidth]{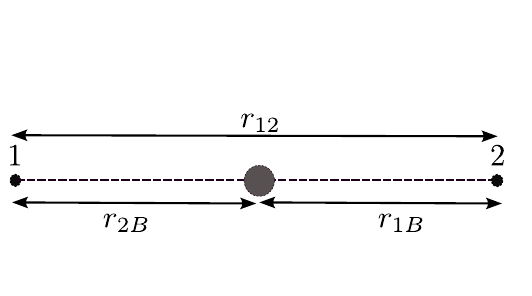}
		\caption{Line-up configuration with a black hole in the middle.}
		\label{fig:BH_UDW_C32}
	\end{subfigure}\hspace{.5cm}
 \begin{subfigure}{.3\textwidth}
		\centering
\includegraphics[width=1.\linewidth]{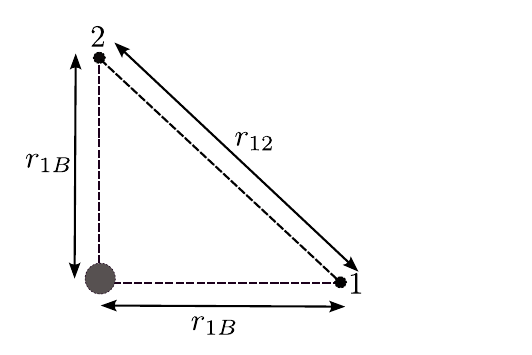}
		\caption{Configuration with a black hole at right-angle vertex.}
		\label{fig:BH_UDW_C33}
	\end{subfigure}
	\caption{Three configurations of UDW probes (small black dots) relative to the black hole (big black dot). }
\label{3_configures}
\end{figure}

We examine three placement configurations of UDW probes relative to the black hole. In configuration (a), illustrated in \cref{fig:BH_UDW_C31}, both UDW probes are positioned along a straight line on the right side of the black hole. The configuration (b), depicted in \cref{fig:BH_UDW_C32}, places the black hole between the two UDW probes. Finally, in configuration (c), shown in \cref{fig:BH_UDW_C33}, the detectors and the black hole form a right triangle. We will also consider some of these configurations when evaluating the quantum discord (section \ref{sec5}) and nonlocality bound (section \ref{sec6}). The placement configuration (a) can be compared with the case considered in \cite{Henderson:2017yuv} using the conventional approach, in which the UDW probes are separated only along the radial direction near a BTZ black hole.

\subsubsection{Configuration a}
The setup illustrated in \cref{fig:BH_UDW_C31} is characterized by two key parameters: the distance $r_{1B}$ between the black hole and the closer of the two UDW probes, and the separation $r_{12}$ between the detectors themselves. We analyze the behavior of concurrence as a function of $T$, $r_{1B}$, and $r_{12}$, keeping all other parameters fixed to the values specified earlier. 

\begin{figure}[ht]
	\centering
	\begin{subfigure}{.4\textwidth}
		\centering
		\includegraphics[width=.9\linewidth]{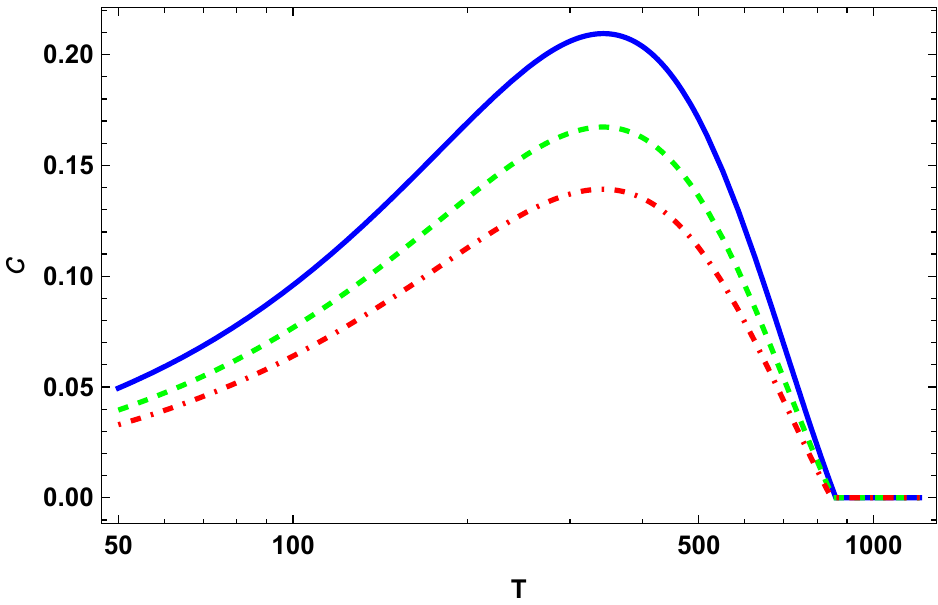}
		\caption{}
		\label{fig:C31_con_vs_T}
	\end{subfigure}\hspace{.45cm}
	\begin{subfigure}{.4\textwidth}
		\centering
		\includegraphics[width=.9\linewidth]{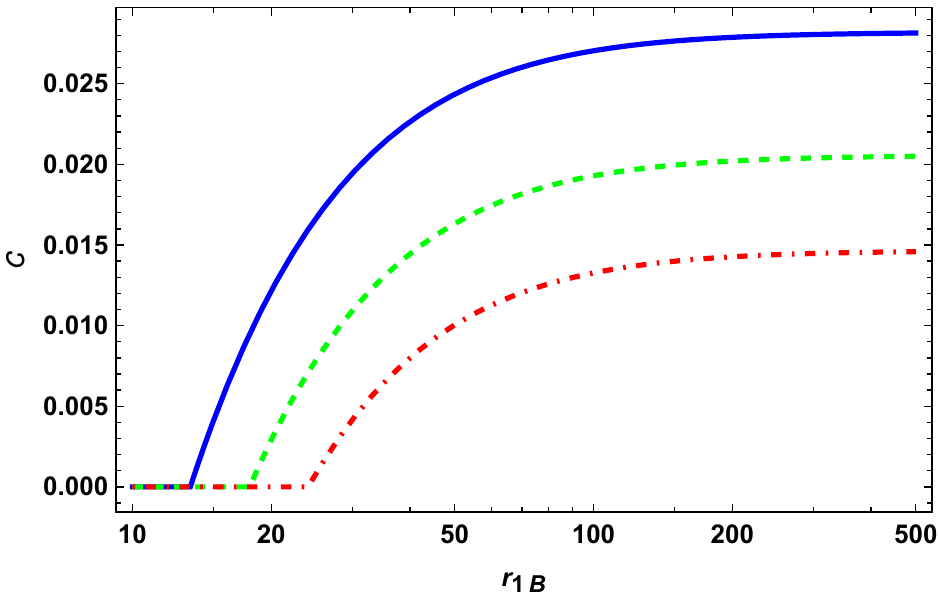}
		\caption{}
		\label{fig:C31_con_vs_r1B}
	\end{subfigure}
 \begin{subfigure}{.4\textwidth}
		\centering
		\includegraphics[width=.9\linewidth]{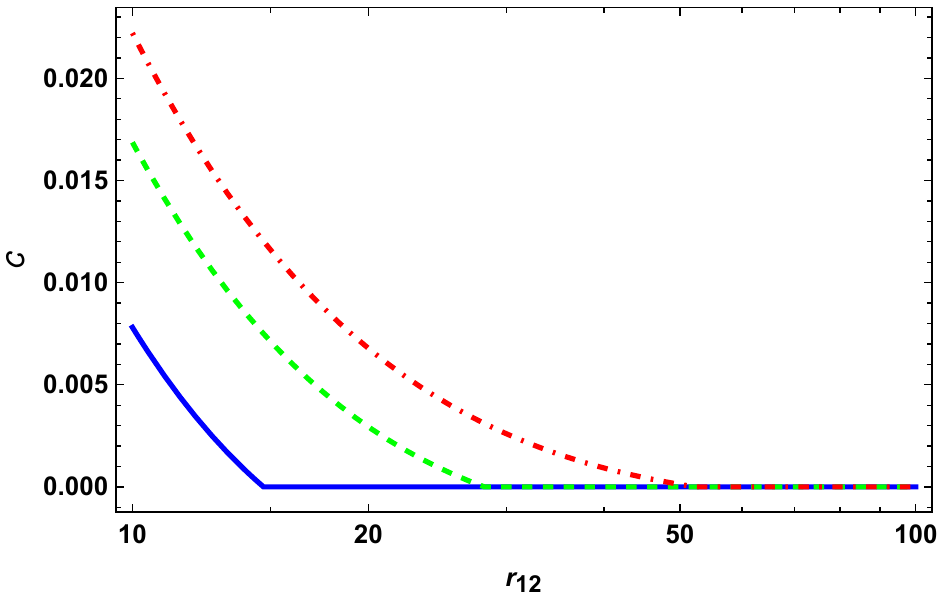}
		\caption{}
		\label{fig:C31_con_vs_r12}
	\end{subfigure}
	\caption{Concurrence $\mathcal{C}$ of configuration \cref{fig:BH_UDW_C31} for $\rho_{12}^f$ of \eq{rho_full}:  (a) $\mathcal{C}$ vs $T$ for  $r_{1B}=10$ and various $r_{12}$: $r_{12}=20$ (solid-blue), $r_{12}=25$ (green-dashed) and $r_{12}=30$ (red-dot-dashed). (b) $\mathcal{C}$ vs  $r_{1B}$ for $r_{12}=25$, and various $T$: $T=900$ (solid-blue), $T=950$ (green-dashed) and $T=1000$ (red-dot-dashed). (c)  $\mathcal{C}$ vs $r_{12}$ for $T=100$, and various $r_{1B}$: $r_{1B}=20$ (solid-blue), $r_{1B}=25$ (green-dashed) and $r_{1B}=30$ (red-dot-dashed).  }
\label{fig_C3_ex1}
\end{figure}

In \cref{fig_C3_ex1}, we present the concurrence for this configuration described by the state $\rho_{12}^f$  \eq{rho_full} as a function of  $T$ in \cref{fig:C31_con_vs_T}, of $r_{1B}$ in \cref{fig:C31_con_vs_r1B} and of $r_{12}$ in \cref{fig:C31_con_vs_r12}.  The behavior of concurrence in  \cref{fig:C31_con_vs_T} resembles the one in \cref{fig_C1_1} with a maximum around $\Omega T \sim {\cal O}(1)$, but with a subtle difference: there exists a ``sudden death" at sufficiently large  $T$ in \cref{fig:C31_con_vs_T}, a feature absent in \cref{fig_C1_1}.  It indicates that the presence of an event horizon can qualitatively alter the patterns of entanglement harvesting.

\FloatBarrier

The ``sudden death" of the concurrence also occurs when $r_{1B}$ is smaller than a critical value, as shown in \cref{fig:C31_con_vs_r1B}, or when $r_{12} $ grows beyond some critical value, as shown in \cref{fig:C31_con_vs_r12}. Note also that the concurrence approaches a constant value as the detectors move farther from the black hole. All of the patterns of entanglement harvesting, as shown in  \cref{fig:C31_con_vs_r1B} and \cref{fig:C31_con_vs_r12} are qualitatively the same as the one observed in \cite{Henderson:2017yuv} by evaluating the entanglement harvesting with the conventional approach based on quantum fields in curved spacetime, and there, the region of vanishing concurrence was coined as ``entanglement shadow". Notably, the size of ``entanglement shadow" increases with the detector's energy gap. These agreements can be seen as a consistency check of the PN-EFT approach. However, in the PN-EFT, we can derive analytical expressions for the reduced states of the UDW probes using the universal transport of the black hole in the low-energy regime. This is beyond the reach of conventional approaches, which require the numerical treatment of the thermal correlators of the mediator fields.

The existence of the ``entanglement shadow" is interesting and requires some explanation. Some possible interpretations may follow. As shown in \cite{Danielson:2022sga, Danielson:2022tdw, Gralla:2023oya, biggs2024comparing}, the cat state of a UDW probe will decrease due to the thermal noise of Hawking radiation. The analog of the cat state for a pair of UDW probes is the entangled state, as the quantum entanglement is a coherent quantum resource. We expect that the strong thermal noise will also destroy the quantum entanglement.  Therefore, when one of the UDW probes moves too close to the black hole, i.e., entering the "entanglement shadow," the two UDW probes cannot become entangled, resulting in "sudden death," as shown in \cref{fig:C31_con_vs_r1B}. Similarly, the coherence for the entanglement cannot be maintained due to the thermal effect if the two UDW probes are separated far enough, as shown in \cref{fig:C31_con_vs_r12}.

\subsubsection{Configuration b}
\begin{figure}[ht]
	\centering
	\begin{subfigure}{.4\textwidth}
		\centering
		\includegraphics[width=.9\linewidth]{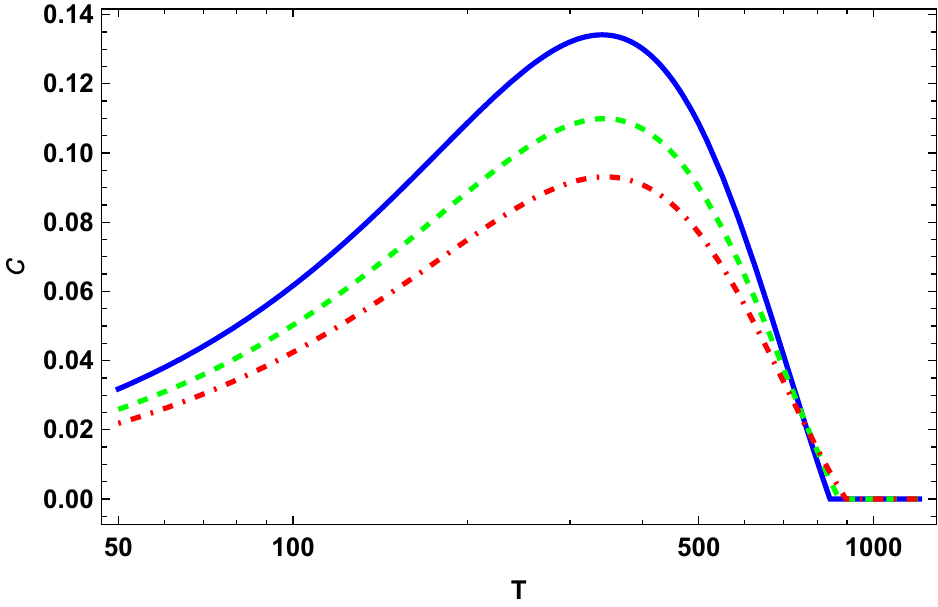}
		\caption{}
		\label{fig:C32_con_vs_T_fix_r1B}
	\end{subfigure}\hspace{.45cm}
	\begin{subfigure}{.4\textwidth}
		\centering
		\includegraphics[width=.9\linewidth]{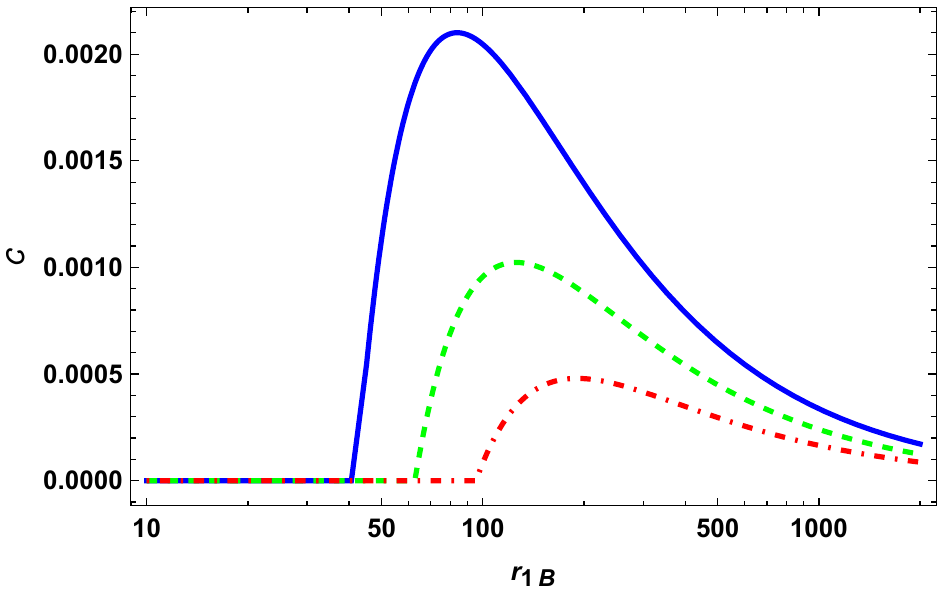}
		\caption{}
		\label{fig:C32_con_vs_r1B_fix_T}
	\end{subfigure}
	\caption{Concurrence $\mathcal{C}$ of configuration \cref{fig:BH_UDW_C32} for $\rho_{12}^f$ of \eq{rho_full}: (a) $\mathcal{C}$ vs $T$ for  various $r_{1B}$: $r_{1B}=20$ (solid-blue), $r_{1B}=25$ (green-dashed) and $r_{2B}=30$ (red-dot-dashed). (b) $\mathcal{C}$ vs  $r_{1B}$ for  various $T$: $T=900$ (solid-blue), $T=950$ (green-dashed) and $T=1000$ (red-dot-dashed).}
   \label{fig_C3_ex2}
\end{figure}

For simplicity, when considering the configuration of \cref{fig:BH_UDW_C32}, we set $r_{1B} = r_{2B}$ so that $r_{12}=2r_{1B}$. Thus, we will consider the concurrence as a function of $T$ and $r_{1B}={r_{12} \over 2}$ by fixing all the other parameters to the values mentioned before. We then plot $\cal C$ vs $T$ for a given $r_{1B}$ in \cref{fig:C32_con_vs_T_fix_r1B}, and $\cal C$ vs $r_{1B}$ for a given $T$ in \cref{fig:C32_con_vs_r1B_fix_T}. 
The pattern shown in  \cref{fig:C32_con_vs_T_fix_r1B} is almost the same as in \cref{fig:C31_con_vs_T}. As in \cref{fig:C31_con_vs_r1B}, there also exists  ``entanglement shadow" in \cref{fig:C32_con_vs_r1B_fix_T}. By comparison, we find that the size of the ``entanglement shadow" for a given $T$ is larger in \cref{fig:C32_con_vs_r1B_fix_T} than in \cref{fig:C31_con_vs_r1B}. This behavior is expected since, in the current configuration, both UDW probes move together toward the black hole so that the thermal noise that causes decoherence is stronger in inhibiting entanglement harvesting than in the configuration of \cref{fig:BH_UDW_C31}. Furthermore, the concurrence no longer saturates to a constant at large $r_{1B}$ but instead decays to zero. This behavior is attributable to the fact that $r_{12} = 2r_{1B}$ in this setup, so increasing $r_{1B}$ also increases the separation between the detectors. Consequently, the behavior of $\mathcal{C}$ at large $r_{12}$ mimics that observed in \cref{fig_C1_2}.

\FloatBarrier

\begin{figure}[ht]
	\centering
	\begin{subfigure}{.4\textwidth}
		\centering
		\includegraphics[width=.9\linewidth]{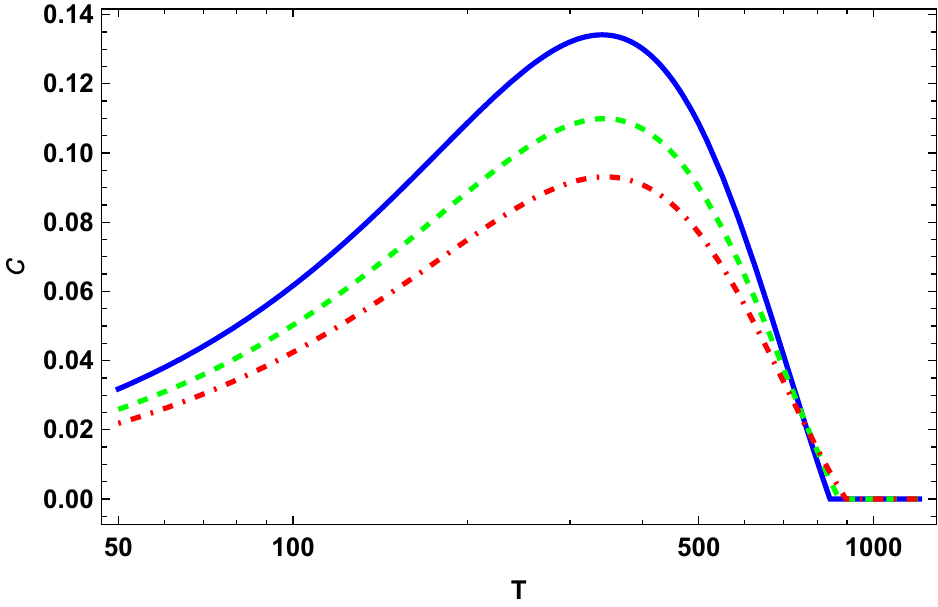}
		\caption{}
		\label{fig:C33_con_vs_T_fix_r1B}
	\end{subfigure}\hspace{.45cm}
	\begin{subfigure}{.4\textwidth}
		\centering
		\includegraphics[width=.9\linewidth]{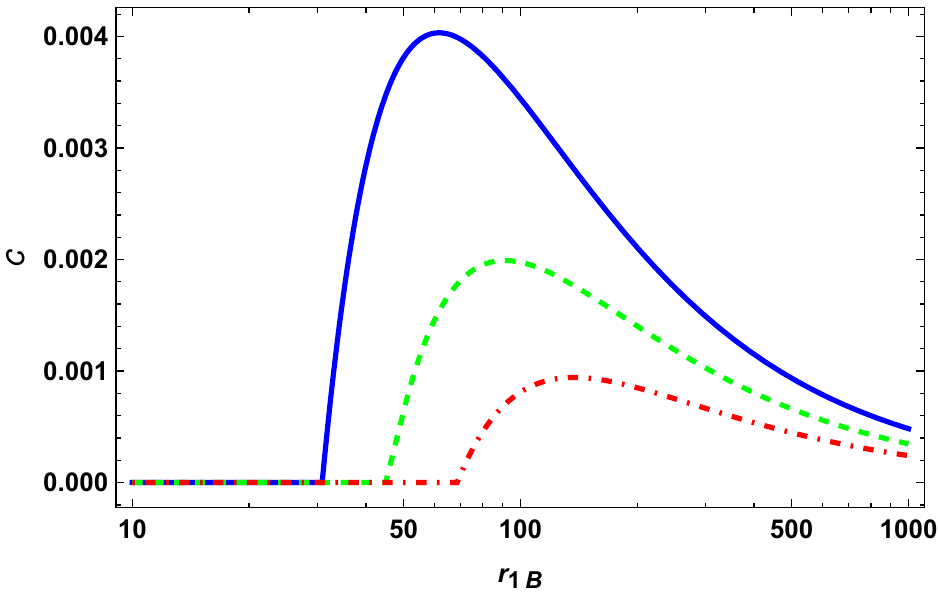}
		\caption{}
		\label{fig:C33_con_vs_r1B_fix_T}
	\end{subfigure}
	\caption{Concurrence $\mathcal{C}$ of configuration \cref{fig:BH_UDW_C33} for $\rho_{12}^f$ of \eq{rho_full}: (a) $\mathcal{C}$ vs $T$ for  various $r_{1B}$: $r_{1B}=20$ (solid-blue), $r_{1B}=25$ (green-dashed) and $r_{2B}=30$ (red-dot-dashed). (b) $\mathcal{C}$ vs  $r_{1B}$ for  various $T$: $T=900$ (solid-blue), $T=950$ (green-dashed) and $T=1000$ (red-dot-dashed). }
\label{fig_C33}
\end{figure}

\subsubsection{Configuration c}
For a given set of $(r_{1B}, r_{2B})$, $r_{12}=r_{1B}+r_{2B}$ for configuration  of \cref{fig:BH_UDW_C32}, and $r_{12}=\sqrt{(r_{1B}^2)+ (r_{2B})^2}$ for configuration of \cref{fig:BH_UDW_C33}. Thus, the difference between configurations b and c will only cause a difference in $X_0$. If we further set $r_{1B}=r_{2B}$, as in configuration b, we expect the results to be qualitatively similar. This is the case when comparing \cref{fig_C3_ex2} and \cref{fig_C33} which consists of $\cal C$ vs $T$ for a given $r_{1B}$ in \cref{fig:C33_con_vs_T_fix_r1B}, and $\cal C$ vs $r_{1B}$ for a given $T$ in \cref{fig:C33_con_vs_r1B_fix_T}. Consequently, in the subsequent analysis of quantum discord and the nonlocality bound, we restrict attention to the configuration shown in \cref{fig:BH_UDW_C32}, omitting further discussion of \cref{fig:BH_UDW_C33}.

\section{Quantum Discord}\label{sec5}

In this section, we examine quantum correlations among the UDW probes, which may not necessarily involve quantum entanglement. Quantum discord is a measure that excludes classical correlations from the total correlations between two subsystems (UDW probes) of a quantum system.  These quantum correlations arise from quantum effects, not necessarily involving entanglement; that is, discord can be nonzero for separable states, which have zero concurrence. Thus, quantum discord can serve as a measure of quantumness other than quantum entanglement, as proposed in \cite{ollivier2001introducing, Henderson:2001wrr}. The notion of ``quantumness," characterized by quantum discord, can be contrasted with the ``quantumness" of QFT, defined by the quantum fluctuations of the quantized fields. The comparison can highlight the contextual difference of ``quantumness".

For a bipartite system composed of two sub-systems $A$ and $B$, the quantum discord is defined by \cite{ollivier2001introducing, Henderson:2001wrr}
\be
{\cal D}(\rho_{AB}) := {\min}_{\{B_k\}} \sum_k p_k S(A\Vert B_k) - S(A\Vert B) \ge 0\;,
\ee
where the relative entropy $S(A\Vert B):=S(AB)-S(B)$ with the von Neumann entropy $S(A):=-{\rm Tr}_A \rho_A \ln \rho_A$, $\{B_k\}$ is a projective measurement basis for performing measurements on subsystem $B$, and $p_k:={\Tr}_B(B_k \rho_{AB})$. The quantum discord vanishes when $\rho_{AB}$ is a pointer state, i.e., $\rho_{AB}=\sum_k B_k \rho_{AB} B_k$. Usually, evaluating quantum discord is tedious for a general $X$-state, as it requires carrying out minimization over possible measurement bases \cite{ali2010quantum, yurischev2015quantum, PhysRevA.88.014302}. However,  in \cite{Lin:2024roh}, we find that the quantum discord for the $X$-state \eq{dtec_den_mat} up to ${\cal O}(g)$ is independent of the measurement basis, and the result is
\bea
{\cal D}(\rho^f_{12}) &=& {1 \over 2 \ln 2} \Bigg[ (P_1 + P_2) \ln (P_1 P_2 - C^2)- 2 P_1 \ln P_1 - 2 P_2 \ln P_2 \ \qquad \nn \\
&& \qquad  + \sqrt{(P_1 -P_2)^2 + 4 C^2} \ln {P_1 + P_2 + \sqrt{(P_1 -P_2)^2 + 4 C^2} \over P_1 + P_2 - \sqrt{(P_1 -P_2)^2 + 4 C^2}} \Bigg] \;.  \qquad \label{def_discord}
\eea
Note that when $C^2=P_1 P_2$, \eq{def_discord} is simplified to
\be
{\cal D}(\rho^f_{12}) = {1 \over 2 \ln 2} \Big[-P_1 \ln P_1-P_2 \ln P_2+(P_1+P_2)\ln(P_1+P_2)\Big]\;.  \qquad \label{def_discord1}
\ee
On the other hand, if $P_1=P_2:=P$,  \eq{def_discord} can be reduced to 
\be \label{QD_id}
{\cal D}(\rho^f_{12}) = {1 \over \ln 2}\Big[ (P+|C|) \ln (P+|C|) + (P-|C|)\log (P-|C|) -2 P \ln P  \Big]\;,
\ee
which can be further reduced to ${\cal D}(\rho^f_{12})=2 P$ if additionally 
$C=P$.

\subsection{Case I}
Since the final state is pure in this case, all bipartite correlations are expected to be inherently quantum. Indeed, the discord is precisely equal to the entanglement entropy of either detector. This can be seen as follows.
For a bipartite pure state, quantum discord is expressed as $\mathcal{D}(A:B)=I(A:B)-J(A:B)$, where the mutual information is $I(A:B)= S(A)+S(B)-S({AB})=2 S(A) \;(\text{or} \; 2S(B))$ and the classical correlation is $J(A:B)=S(A)$. Consequently, the quantum discord reduces to $\mathcal{D}(A:B)=S(A)\;(\text{or} \;S(B))$.

\subsection{Case II and III}
As discussed before, for the $X$-states up to ${\cal O}(g)$, the quantum discord is independent of $(\rho^f_{12})_{14}=(\rho^f_{12})_{41}$, i.e., $X$ or $X_0$, thus it will be the same for Case II and Case III even though only the latter can have nonzero concurrence. Furthermore, using \eq{C2=PP}, we notice that the quantum discord for both cases is given by \eq{def_discord1}. 

In the following, we present the corresponding quantum discord for the configurations of  \cref{fig:BH_UDW_C31,fig:BH_UDW_C32}, respectively.

\subsubsection{Configuration a}
\begin{figure}[ht]
	\centering
	\begin{subfigure}{.4\textwidth}
		\centering
		\includegraphics[width=.9\linewidth]{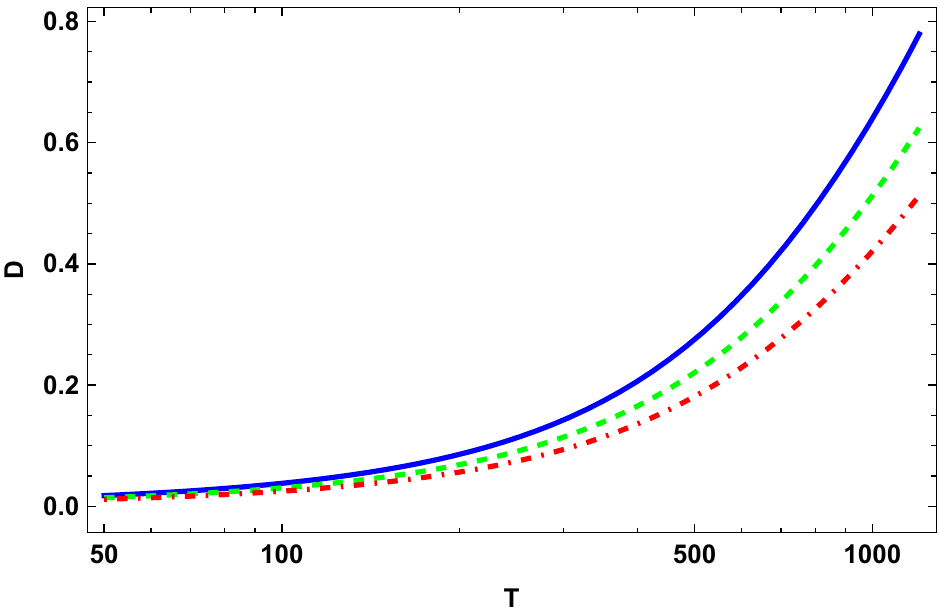}
		\caption{}
		\label{fig:QD21_vs_T}
	\end{subfigure}\hspace{.45cm}
	\begin{subfigure}{.4\textwidth}
		\centering
		\includegraphics[width=.9\linewidth]{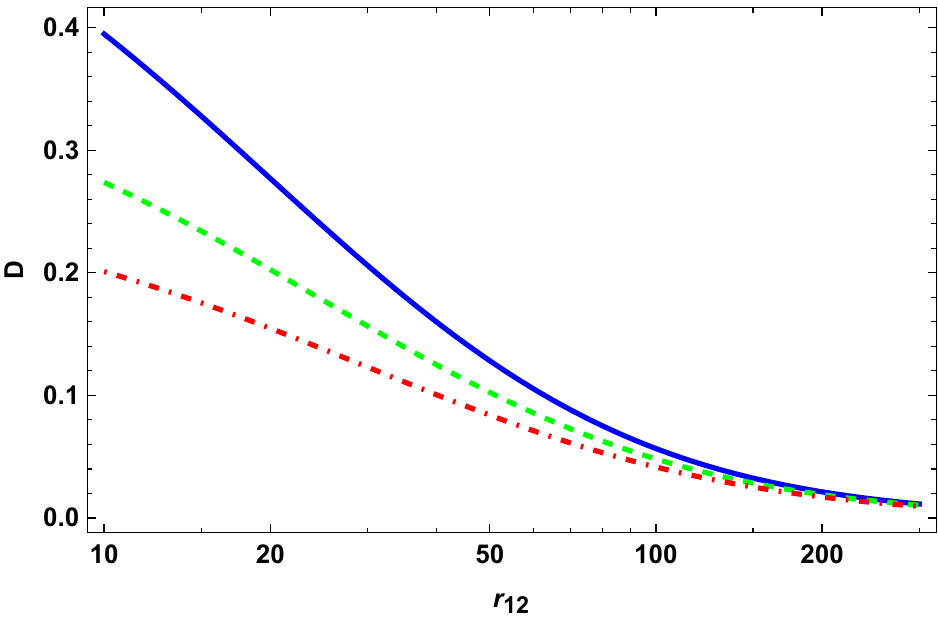}
		\caption{}
		\label{fig:QD21_vs_r12}
	\end{subfigure}
 \begin{subfigure}{.4\textwidth}
		\centering
		\includegraphics[width=.9\linewidth]{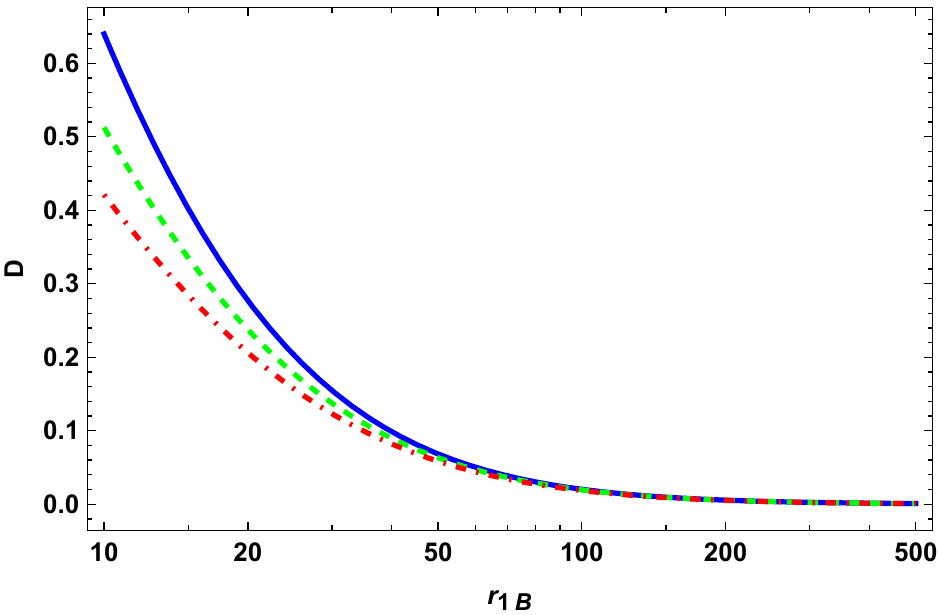}
		\caption{}
		\label{fig:QD21_vs_r1B}
	\end{subfigure}
	\caption{Quantum discord $D$ of configuration \cref{fig:BH_UDW_C31} for $\rho_{12}^f$ of \eq{rho_full}:  (a) $D$ vs $T$ for $r_{1B}=10$ and various $r_{12}$: $r_{12}=20$ (solid-blue), $r_{12}=25$ (green-dashed) and $r_{12}=30$ (red-dot-dashed). (b) $D$ vs  $r_{1B}$ for $T=1000$, and various $r_{12}$: $r_{12}=20$ (solid-blue), $r_{12}=25$ (green-dashed) and $r_{12}=30$ (red-dot-dashed). (c)  $D$ vs $r_{12}$ for $T=100$, and various $r_{1B}$: $r_{1B}=20$ (solid-blue), $r_{1B}=25$ (green-dashed) and $r_{1B}=30$ (red-dot-dashed).  }
\label{fig:QD21}
\end{figure}

The results of quantum discord for the configuration of \cref{fig:BH_UDW_C31} obtained by \eq{def_discord1} are presented in \cref{fig:QD21} with ${\cal D}$ as a function of $T$ in \cref{fig:QD21_vs_T}, of $r_{12}$ in \cref{fig:QD21_vs_r12}, and of $r_{1B}$ in \cref{fig:QD21_vs_r1B}  by fixing  rest of the parameters. 

\FloatBarrier

We observe that the results for quantum discord are consistent with expectations. Specifically, the pure quantum correlation increases monotonically with the interaction time and decreases monotonically with the interdistance between UDW probes. Unlike the bump behavior of concurrence shown in \cref{fig:C32_con_vs_r1B_fix_T}, quantum discord decreases monotonically as the UDW probes with a fixed separation move away from the black hole. Besides, the key feature distinguishing the concurrence and quantum discord is the absence of ``sudden death" behavior for the latter. This implies that the UDW probes are quantum-correlated with the black hole, and the thermal noise of Hawking radiation will not affect the pure quantum correlations but may affect the classical ones. Furthermore, from \cref{fig:QD21_vs_r12} and \cref{fig:QD21_vs_r1B}, we find that discord decays more rapidly with $r_{1B}$ than with $r_{12}$.

\subsubsection{Configuration b}

\begin{figure}[ht]
	\centering
	\begin{subfigure}{.4\textwidth}
		\centering
		\includegraphics[width=.9\linewidth]{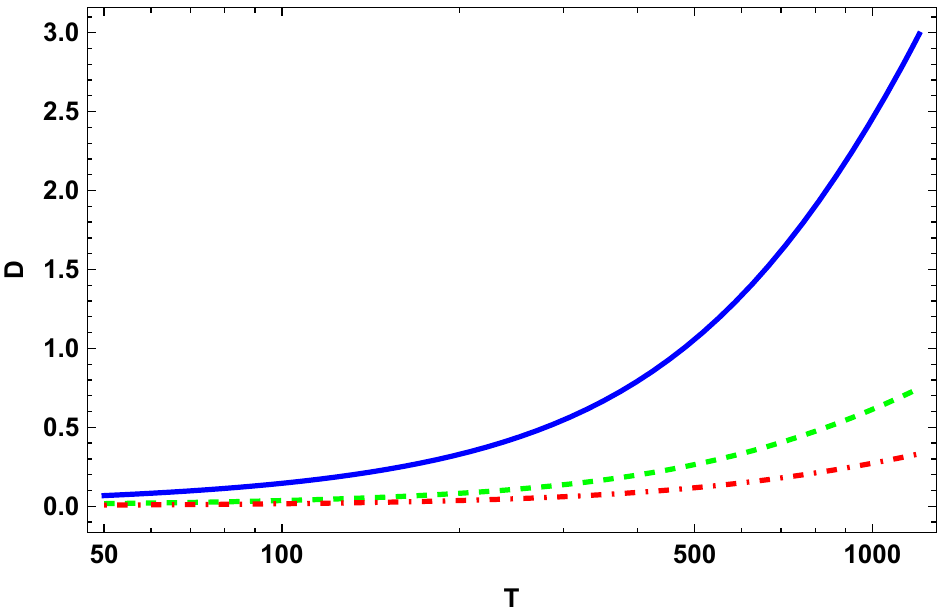}
		\caption{}
		\label{fig:QD22_vs_T}
	\end{subfigure}\hspace{.45cm}
 \begin{subfigure}{.4\textwidth}
		\centering
		\includegraphics[width=.9\linewidth]{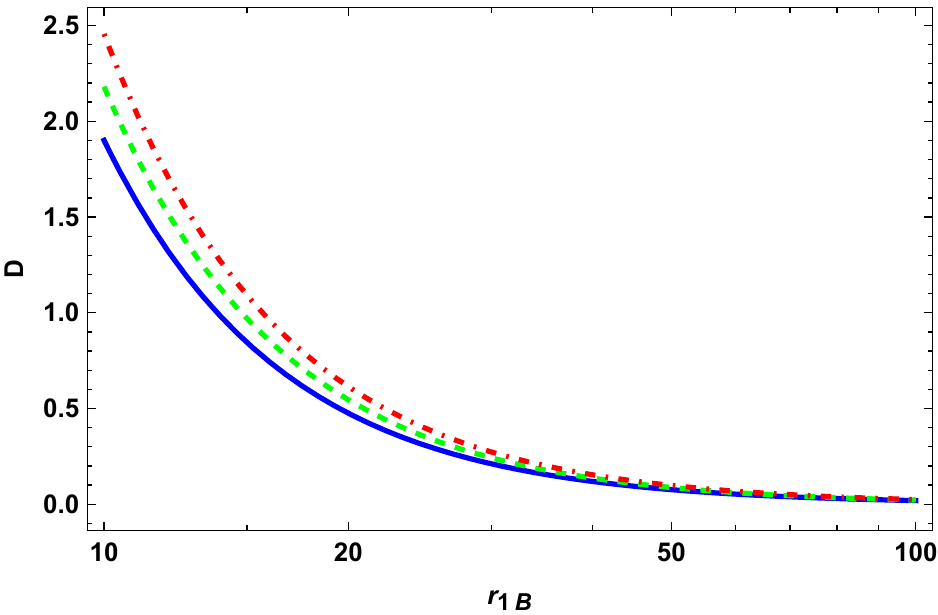}
		\caption{}
		\label{fig:QD22_vs_r1B}
	\end{subfigure}
	\caption{ Quantum discord $D$ of configuration \cref{fig:BH_UDW_C32} for $\rho_{12}^f$ of \eq{rho_full}: (a) $D$ vs $T$ for  various $r_{1B}$: $r_{1B}=20$ (solid-blue), $r_{1B}=25$ (green-dashed) and $r_{1B}=30$ (red-dot-dashed). (b) $D$ vs  $r_{1B}$ for  various $T$: $T=800$ (solid-blue), $T=900$ (green-dashed) and $T=1000$ (red-dot-dashed). }
\label{fig:QD22}
\end{figure}

The results of quantum discord for the configuration of \cref{fig:BH_UDW_C32} with $r_{1B}=r_{2B}=r_{12}/2$ are presented in \cref{fig:QD22}: $\cal D$ vs $T$ for various $r_{1B}$ in \cref{fig:QD22_vs_T}, and $\cal D$ vs $r_{1B}$ for various $T$ in \cref{fig:QD22_vs_r1B}. The patterns found here are similar to those of \cref{fig:QD21} and bear analogous implications.

\FloatBarrier

\section{Nonlocality Bound}\label{sec6}

The EPR paper \cite{PhysRev.47.777} raised the issue of the nonlocal nature of the quantum entangled states, now recognized as the quantum resource \cite{buhrman2010nonlocality,  Chitambar:2018rnj} to achieve nontrivial tasks such as quantum teleportation or dense coding \cite{Bennett:1992tv, Bennett:1992zzb}. In a broad context, nonlocality is examined through Bell-type experiments, which analyze correlations in measurement outcomes in a bipartite communication channel and derive algebraic inequalities that characterize the nonlocal resource of the channel. If there is no nonlocal resource, then the joint probability of the bipartite measurement outcomes takes a form admitted by the local hidden variable model (LHVM)
\be\label{LHVM}
p(ab|xy)=\int_{\Lambda} d\lambda q(\lambda) p(a|x,\lambda) p(b|y,\lambda)\;,
\ee
where $x$ and $a$ are respectively Alice's measurement choice and outcome, and $y$ and $b$ are Bob's ones, and $q(\lambda)$ is the probability distribution of local hidden variable $\lambda$. On the other hand, in quantum mechanics, the joint probability is constructed by
\be
p(ab|xy) = {\rm tr} (\rho_{AB} M_{a|x} M_{b|y})\;,
\ee
where $\{ M_{a|x} \}$ are the positive operator-valued measure (POVM). Then, there exist some nonlocality measures characterized by the following kinds of inequalities \cite{Brunner:2013est}
\be
\sum_{\{abxy\}} s^k_{abxy} \; p(ab|xy) \le S_k\;. 
\ee
Many such inequalities labeled by $k$ may exist. If one of the above inequalities is violated, nonlocal resources will exist in the communication channel.

We will only consider bipartite communication tasks with two-level systems, i.e., $a,b,x,y=\{-1,+1\}$. In such a case, the inequality characterizing the nonlocality is the  Clauser-Horne-Shimony-Holt (CHSH) inequality
\be\label{CHSH}
S:=\langle a_{-1} b_{-1} \rangle + \langle a_{-1} b_{+1} \rangle + \langle a_{+1} b_{-1} \rangle - \langle a_{+1} b_{+1} \rangle \le 2\;,
\ee
with $\langle a_x b_y \rangle = \sum_{\{ab\}} ab \; p(ab|xy)$. Thus, the CHSH nonlocality bound is $S_{k={\rm CHSH}}=2$. One can also extend the above inequality study from nonlocality to no-signaling \cite{1980LMaPh...4...93C, Khalfin1992-KHAQCI, Rohrlich:1995cf, Brunner:2013est}, i.e., with the no-signaling conditions: $p(a|x)=p(a|xy):=\sum_{\{b\}} p(ab|xy)$ and $p(b|y)=p(b|xy):=\sum_{\{a\}} p(ab|xy)$. The no-signaling inequality for the bipartite qubits' {\it quantum} communication turns out to be the same as \eq{CHSH} but replacing $S_{\rm CHSH}=2$ with the Tsirelson bound $S_{\rm Tsirelson} = 2\sqrt{2}$.

In this paper, we will adopt the reduced final state $\rho^f_{12}$ of UDW qubits as the communication channel for CHSH inequality and then examine its (non)locality. In this case, if a quantum state is separable, i.e., $\rho_{A B}=\sum_\lambda p_\lambda \rho_A^\lambda \otimes \rho_B^\lambda$, its corresponding $p(ab|xy)$ can be put into the form of \eq{LHVM}, thus it is local. According to the example considered in the EPR paper, one may infer that all entangled quantum states are nonlocal. In \cite{Capasso1,GISIN1991201,Selleri1}, it has been proven that all entangled pure states are nonlocal. However, this is not the case for entangled mixed states; for example, some Werner states \cite{Werner} are local. For a quantum state $\rho$ of bipartite qubits, the quantity $S$ in the CHSH inequality \eq{CHSH}, denoting it  by $S_{\rho}$ since it is evaluated for a quantum state, is given by \cite{HORODECKI1995340}
\be\label{CHSH_violation}
S_\rho=2 \sqrt{t_{11}^2+t_{22}^2}\;,
\ee
where  $t_{11}^2$ and $t_{22}^2$ are the two largest eigenvalues of the matrix $T_\rho T_\rho^T$ with the $3\times 3$ matrix $(T_{\rho})_{ij}:={\mathrm {tr}}\big[\rho (\sigma_i \otimes \sigma_j) \big]$ for $i,j=1,2,3$.
If $S_{\rho} > S_{\mathrm {CHSH}}=2$, then the state $\rho$ can be utilized as a nonlocal quantum resource. Unlike concurrence or negativity,  $S_{\rho}$ is not an entanglement measure as it is not a monotone under LOCC  \cite{PhysRevLett.89.170401}. Thus, it is possible to have local entangled mixed states, such as will be seen for the $\rho^f_{12}$ considered in this paper. Below, we will evaluate $S_{\rho}$ up to $\mathcal{O}(g)$ to be consistent with the perturbation theory and show its behaviors for Cases I, II, and III.

\subsection{Case I}
For this case, we find that $S_{\rho} = S_{\mathrm {CHSH}} = 2$, indicating that the CHSH inequality is saturated but not violated for any values of $T$ and $r_{12}$. It implies that the reduced final state $\rho^f_{12}$ \eq{red_final_noBH} obtained from evolution by Coulombic interaction is entangled but does not violate the CHSH inequality. The local nature of the correlations in Bell-type experiments contrasts sharply with the inherently nonlocal and acausal nature of Coulombic forces in a field-theoretic context. 

\subsection{Case II}
 We now examine the nonlocality bound for Case II with a black hole, but not the mutual Coulombic interaction between UDW probes. We will only consider the configurations of  \cref{fig:BH_UDW_C31,fig:BH_UDW_C32}.

\subsubsection{Configuration a}
\begin{figure}[ht]
	\centering
	\begin{subfigure}{.4\textwidth}
		\centering
		\includegraphics[width=.9\linewidth]{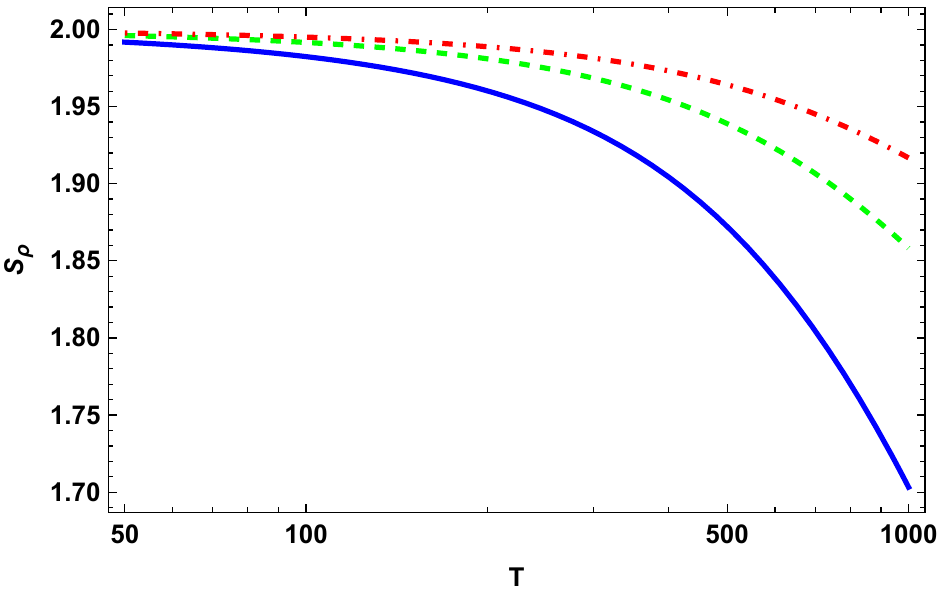}
		\caption{}
		\label{fig:NL21_vs_T}
	\end{subfigure}\hspace{.45cm}
	\begin{subfigure}{.4\textwidth}
		\centering
		\includegraphics[width=.9\linewidth]{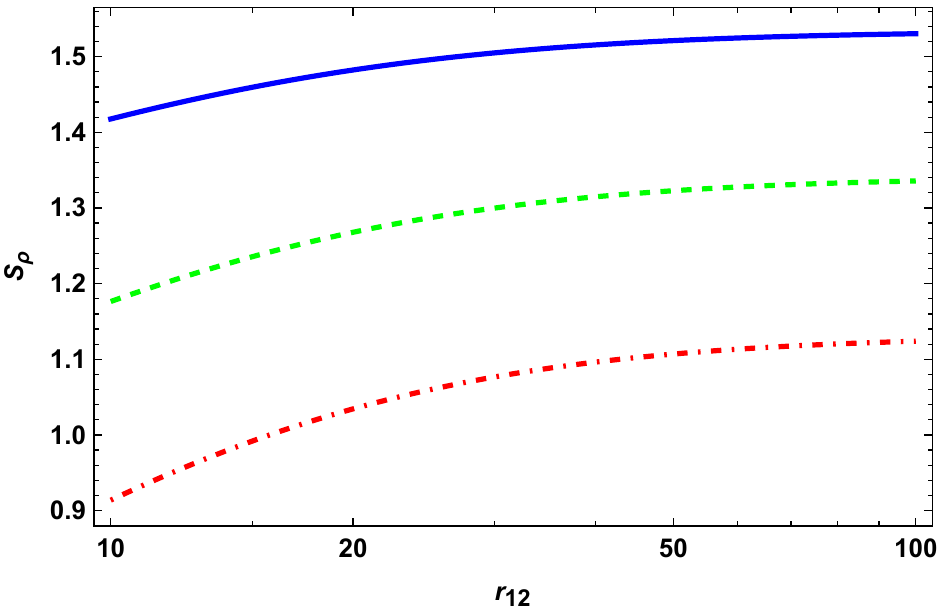}
		\caption{}
		\label{fig:NL21_vs_r12}
	\end{subfigure}
 \begin{subfigure}{.4\textwidth}
		\centering
		\includegraphics[width=.9\linewidth]{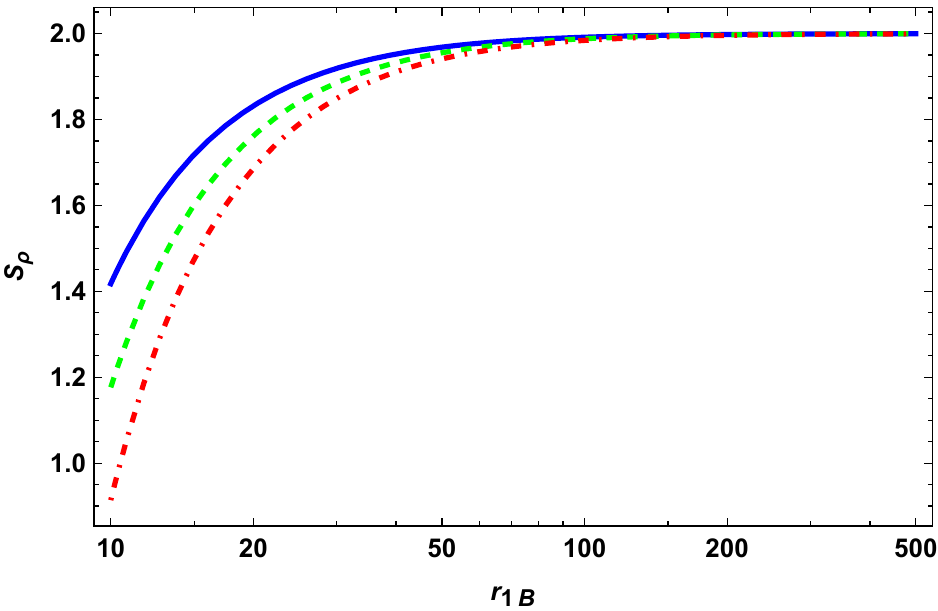}
		\caption{}
		\label{fig:NL21_vs_r1B}
	\end{subfigure}
	\caption{Quantum nonlocality  $S_{\rho}$ for $\rho_{12}^f$ of \eq{dtec_den_mat}:  (a) $S_{\rho}$ vs $T$ for $r_{12}=20$ and various $r_{1B}$: $r_{1B}=50$ (solid-blue), $r_{1B}=75$ (green-dashed) and $r_{1B}=100$ (red-dot-dashed). (b) $S_{\rho}$ vs  $r_{12}$ for   $r_{1B}=10$ and various $T$: $T=150$ (solid-blue), $T=200$ (green-dashed) and $T=250$ (red-dot-dashed). (c)  $S_{\rho}$ vs  $r_{1B}$ for  $r_{12}=10$ and various $T$: $T=150$ (solid-blue), $T=200$ (green-dashed) and $T=250$ (red-dot-dashed). }
		\label{fig:NL21}
\end{figure}

In \cref{fig:NL21_vs_T}, we plot $S_{\rho}$ as a function of $T$, where we see a monotonic decreasing behavior of  $S_{\rho}$. The nonlocality bound is not violated for any value of $T$. Furthermore, we present $S_{\rho}$ in \cref{fig:NL21_vs_r12,fig:NL21_vs_r1B} with $r_{12}$ and $r_{1B}$ respectively, where we notice  $S_{\rho}$ increases monotonically as $r_{12}$ and $r_{1B}$ grows. Again, the nonlocality bound is not violated for any value of $r_{12}$ and $r_{1B}$. As $S_{\rho}$ is not a monotone measure under LOCC, it is unclear if these monotonic behaviors can be interpreted as the degrees of nonlocality of the underlying quantum state. Despite that, these patterns are some physical characteristics of these quantum states.

\FloatBarrier
\FloatBarrier

\subsubsection{Configuration b}
\begin{figure}[ht]
	\centering
	\begin{subfigure}{.4\textwidth}
		\centering
		\includegraphics[width=.9\linewidth]{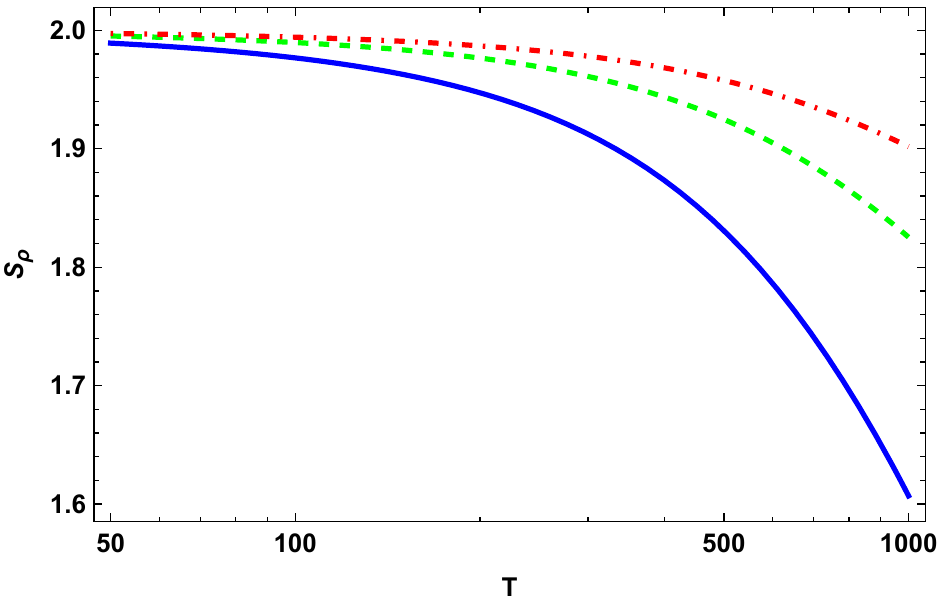}
		\caption{}
		\label{fig:NL22_vs_T}
	\end{subfigure}\hspace{.45cm}
	\begin{subfigure}{.4\textwidth}
		\centering
		\includegraphics[width=.9\linewidth]{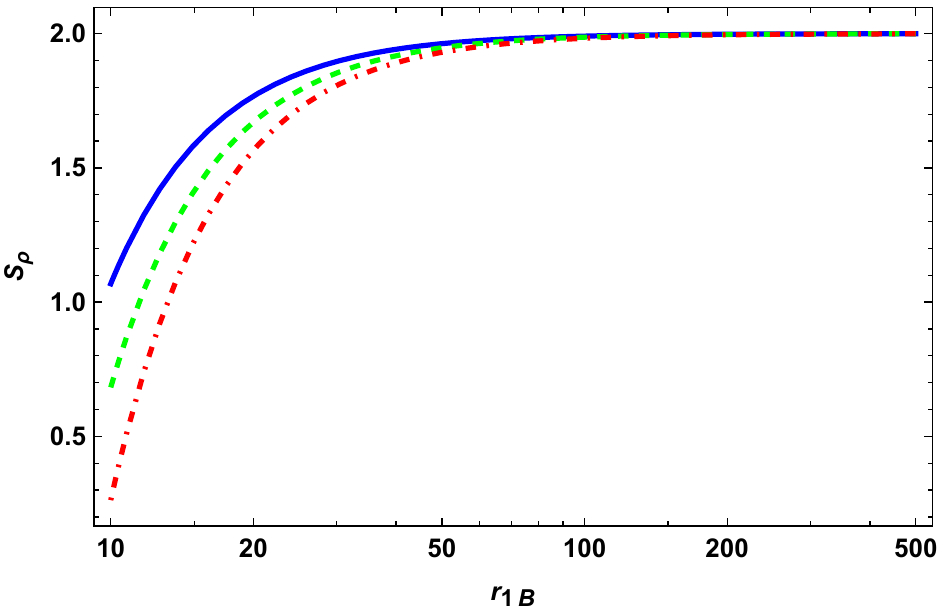}
		\caption{}
		\label{fig:NL22_vs_r1B}
	\end{subfigure}
	\caption{Quantum nonlocality  $S_{\rho}$ for $\rho_{12}^f$ of \eq{dtec_den_mat}:  (a) $S_{\rho}$ vs $T$ for various $r_{1B}$: $r_{1B}=50$ (solid-blue), $r_{1B}=75$ (green-dashed) and $r_{1B}=100$ (red-dot-dashed).  (b) $S_{\rho}$ vs  $r_{1B}$ for various $T$: $T=150$ (solid-blue), $T=200$ (green-dashed) and $T=250$ (red-dot-dashed). }
\label{fig:NL22}
\end{figure}

We present $S_{\rho}$ as a function of $T$ and $r_{1B}$ in \cref{fig:NL22_vs_T,fig:NL22_vs_r1B} respectively. The patterns shown here are qualitatively the same as in 
\cref{fig:NL21_vs_T,fig:NL21_vs_r1B}. In particular, the CHSH nonlocality bound is not violated for all the parameter ranges considered. It implies that the quantum locality of Case II is configuration-independent. 

\FloatBarrier
\FloatBarrier

\begin{figure}[ht]
	\centering
	\begin{subfigure}{.4\textwidth}
		\centering
		\includegraphics[width=.9\linewidth]{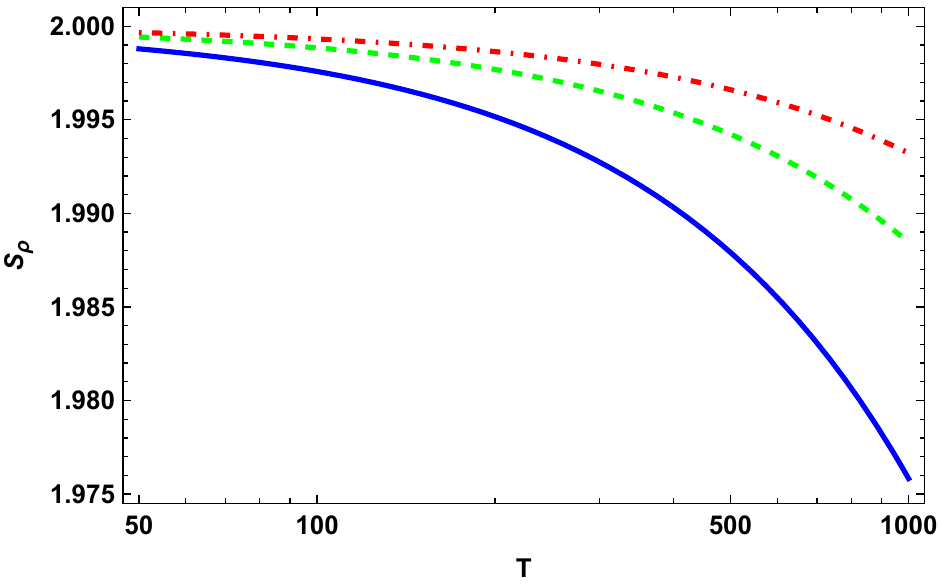}
		\caption{}
		\label{fig:NL31_vs_T}
	\end{subfigure}\hspace{.45cm}
	\begin{subfigure}{.4\textwidth}
		\centering
		\includegraphics[width=.9\linewidth]{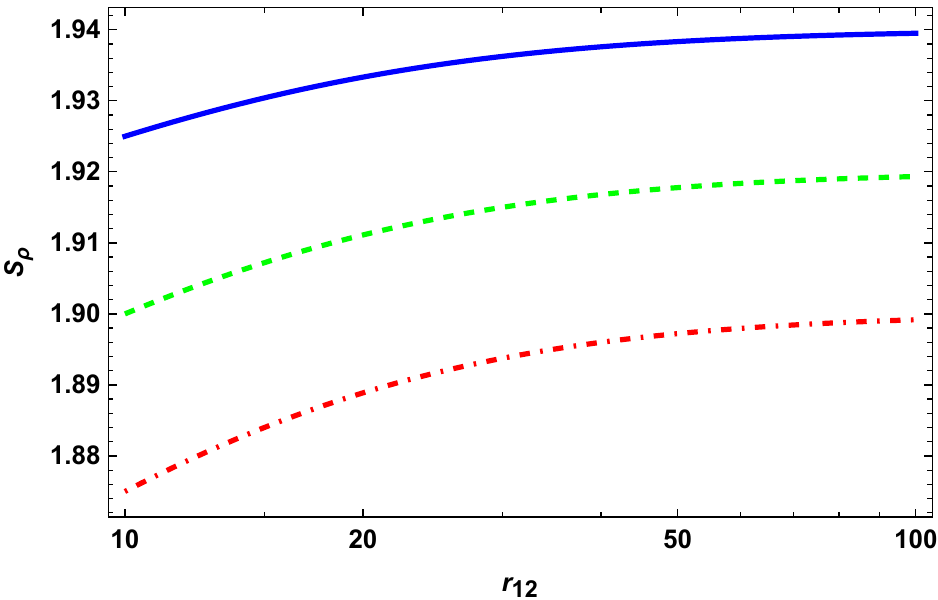}
		\caption{}
		\label{fig:NL31_vs_r12}
	\end{subfigure}
 \begin{subfigure}{.4\textwidth}
		\centering
		\includegraphics[width=.9\linewidth]{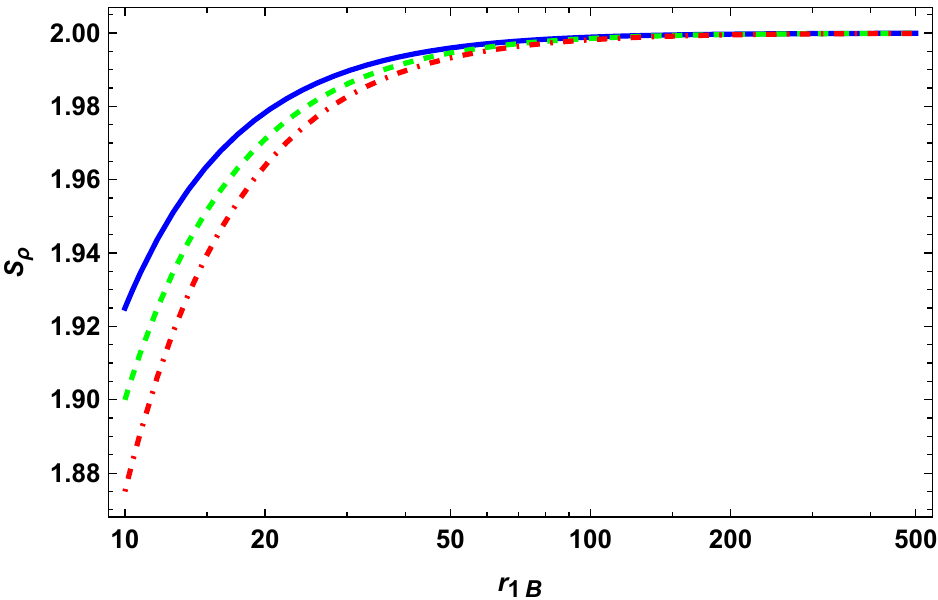}
		\caption{}
		\label{fig:NL31_vs_r1B}
	\end{subfigure}
	\caption{ Quantum nonlocality $S_{\rho}$ for $\rho_{12}^f$ of \eq{rho_full}:  (a) $S_{\rho}$ vs $T$ for $r_{12}=20$ and various $r_{1B}$: $r_{1B}=50$ (solid-blue), $r_{1B}=75$ (green-dashed) and $r_{1B}=100$ (red-dot-dashed). (b) $S_{\rho}$ vs  $r_{12}$ for $r_{1B}=10$ and various $T$: $T=150$ (solid-blue), $T=200$ (green-dashed) and $T=250$ (red-dot-dashed). (c)  $S_{\rho}$ vs  $r_{1B}$ for  $r_{12}=10$ and various $T$: $T=150$ (solid-blue), $T=200$ (green-dashed) and $T=250$ (red-dot-dashed). }
		\label{fig:NL31}
\end{figure}
%

\subsection{Case III}
This subsection considers the case by including all the EFT's interactions. Again, $S_{\rho}$ of the configurations of  \cref{fig:BH_UDW_C31,fig:BH_UDW_C32} will be presented.

\subsubsection{Configuration a}
We first plot $S_{\rho}$ as a function of $T$ in \cref{fig:NL31_vs_T}, where it exhibits a monotonic decreasing behavior. Furthermore, in \cref{fig:NL31_vs_r12,fig:NL31_vs_r1B} we plot $S_{\rho}$ with $r_{12}$ and $r_{1B}$ respectively. We see no violation of the nonlocality bound in all three plots, and the patterns are qualitatively the same as in Case II.

\subsubsection{Configuration b}
\begin{figure}[ht]
 	\centering
 	\begin{subfigure}{.4\textwidth}
 		\centering
 		\includegraphics[width=.9\linewidth]{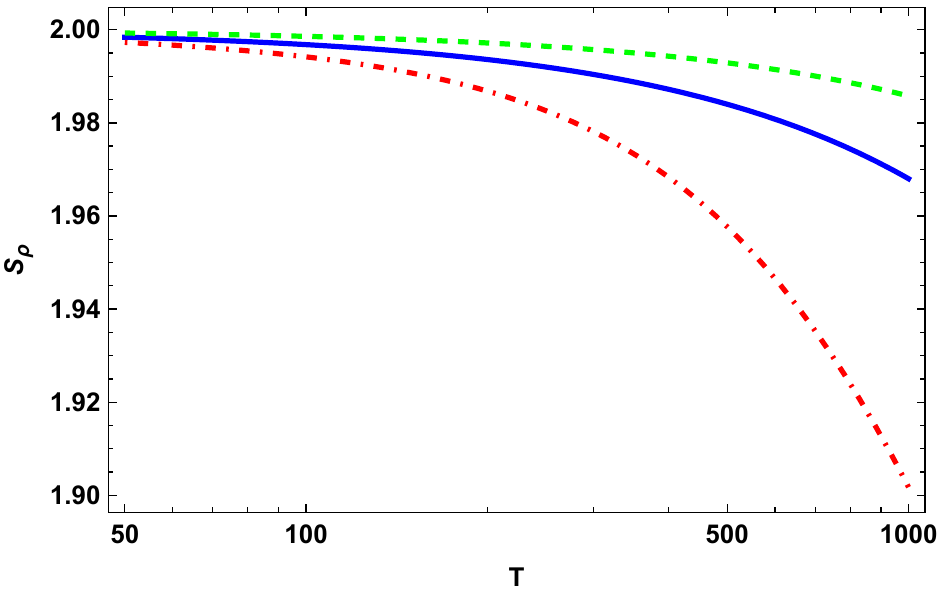}
		\caption{}
		\label{fig:NL32_vs_T}
 	\end{subfigure}\hspace{.45cm}
 	\begin{subfigure}{.4\textwidth}
 		\centering
 		\includegraphics[width=.9\linewidth]{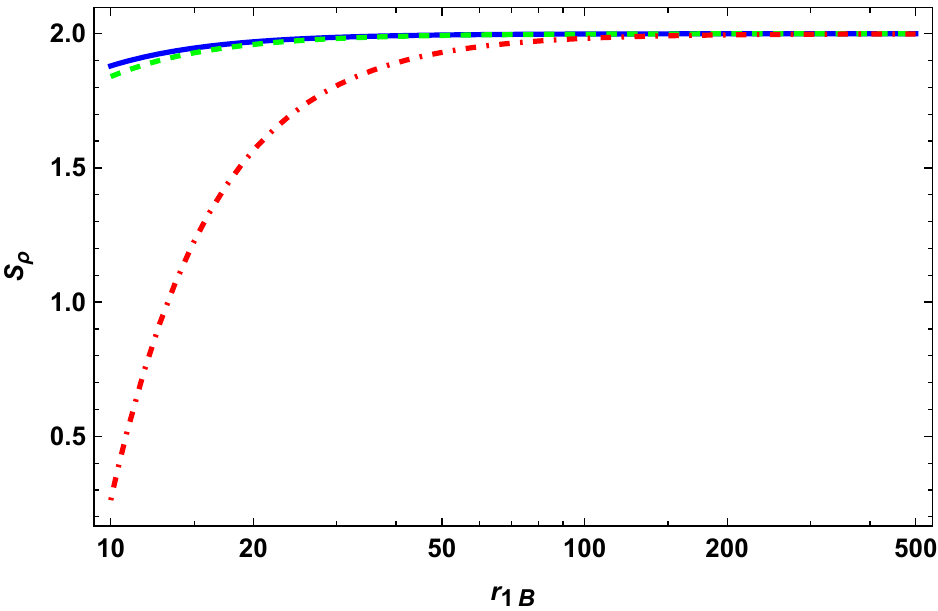}
 		\caption{}
 		\label{fig:NL32_vs_r1B}
 	\end{subfigure}
 	\caption{Quantum nonlocality  $S_{\rho}$ for $\rho_{12}^f$ of \eq{rho_full}:  (a) $S_{\rho}$ vs $T$ for various $r_{1B}$ $r_{1B}=50$ (solid-blue), $r_{1B}=75$ (green-dashed) and $r_{1B}=100$ (red-dot-dashed).  (b) $S_{\rho}$ vs  $r_{1B}$ for various $T$: $T=150$ (solid-blue), $T=200$ (green-dashed) and $T=250$ (red-dot-dashed). }
 \label{fig:NL32}
 \end{figure}

Here we plot $S_{\rho}$ with $T$ and $r_{1B}$ in \cref{fig:NL32_vs_T,fig:NL32_vs_r1B} respectively. We notice the pattern of the nonlocality bound analogous to previously observed: in particular, $S_{\rho}$ is always bounded by $S_{\rm CHSH}=2$. It again indicates the local nature of the correlations. Thus, the quantum state in Cases II and III is local in Bell's sense but can produce nonclassical correlations because quantum discord is nonvanishing.

%

\section{Conclusion and Discussion}\label{sec7}

In this paper, we investigated the extraction of quantum information from a quantum black hole using a pair of UDW detectors as probes. These results shed light on the structure of quantum information hidden behind the event horizon. Instead of using the conventional approach based on quantum field theory in curved spacetime, we adopted a new framework that treats black holes as tidally deformable thermal bodies under the dynamical quantum fluctuations of the surrounding massless mediator fields. Based on this, we have developed a post-Newtonian-type effective field theory (PN-EFT) that describes the interactions between the black hole's tidally induced multipole moments and the intrinsic multipole moments of the UDW probes. The black holes' multipole moments were derived by combining dynamical tidal effects with thermal Hawking effects. They can be regarded as the quantum degrees of freedom of black holes. More importantly, the low-energy behavior of their Wightman functions is universal, partly due to the no-hair theorem. Using this simple universality, we derived the analytical form of the reduced states of the UDW probes from PN-EFT.

From the analytically reduced states, we obtained three quantum information measures: entanglement harvesting, quantum discord characterizing pure quantum correlations, and Bell's nonlocality bound—in a straightforward manner, without encountering the delicate thermal correlators of the mediator fields, as in the conventional approach. This may be part of the reason why quantum information measures around a black hole, other than entanglement harvesting, have not been considered previously. By properly tuning the relative interaction strengths in the leading-order action of PN-EFT, we have considered three cases: (I) without the black hole, (II) without the direct mutual interaction between UDW probes, and (III) with both. We summarize the qualitative results for all three quantum information measures in all three cases in \cref{tab1}.

\begin{table}[ht]
    \centering
    \begin{tabular}{cccc}
    \hline\hline
& \raisebox{-1ex}{Case I} & \raisebox{-1ex}{Case II} & \raisebox{-1ex}{Case III} \\ [1.5ex]
         \hline
         \raisebox{-1ex}{Entanglement} & \raisebox{-2ex}{Yes}  &  \raisebox{-2ex}{No} &  \raisebox{-2ex}{Yes}   \\ [1.5ex] Harvesting   \\
      \hline
     \raisebox{-1ex}{Quantum Correlaiton}  & \raisebox{-1ex}{Yes} & \raisebox{-1ex}{Yes} & \raisebox{-1ex}{Yes} \\ [1.5ex]
     \hline
      \raisebox{-1ex}{Bell's Nonlocality}  & \raisebox{-1ex}{No}  & \raisebox{-1ex}{No} & \raisebox{-1ex}{No} \\ [1.5ex]
     \hline
    \end{tabular}
    \caption{Summary of three quantum information measures of the UDW probes, which characterize the entanglement harvesting, pure quantum correlations, and Bell's nonlocality, for the three cases: (I) without black hole; (II) without the direct mutual interaction between UDW probes; (III) with both. By comparing the items in each row of the table, we recognize the relevance of various interactions of the PN-EFT to the corresponding quantum information measure. }
    \label{tab1}
\end{table}

From \cref{tab1}, we can see that in all three cases, the nonlocality bound is not violated.  Furthermore, the detailed plots in \cref{sec6} reveal that the nonlocality measure exhibits qualitatively similar patterns across all scenarios. This suggests that, within our framework, quantum black holes do not violate the nonlocality bound. Regarding entanglement harvesting, our results indicate that mutual interaction among the Unruh-DeWitt (UDW) detectors is essential, whereas the presence of the black hole is not. The reason is that the mediator fields are quantized; thus, by themselves, they carry quantum information that entangles the two detector systems. One can find a more thorough discussion in \cite{Lin:2025ipk}.
On the other hand, thermal noise from black holes can counteract the quantum nature of the contribution to entanglement harvesting. This comparison demonstrates the advantage of our Post-Newtonian Effective Field Theory (PN-EFT) approach over the conventional approach, in which such comparisons cannot be made directly because the black hole is treated as a background spacetime. We also note that in case II, non-zero quantum correlations persist even in the absence of entanglement harvesting. If these features of the quantum information measures are interpreted as arising from communication with the quantum black hole, they may reflect the intrinsic properties of the information content within the black hole.

The results presented herein are based on the leading-order action of the PN-EFT. Further investigation that incorporates higher-order PN interactions, which account for radiation and backreaction effects, is therefore a promising direction for future work. As a precursor study, this work aims to open a new avenue for exploring quantum black holes with novel tools and perspectives.

\section*{Acknowledgment}
We thank Huan-Yu Ku for the helpful discussions. SM is grateful to Ignacio Araya and Giorgos Anastasiou for their hospitality at Universidad Andrés Bello and Universidad Adolfo Ibañez, respectively. The work of FLL is supported by Taiwan's National Science and Technology Council (NSTC) with grant number 112-2112-M-003-006-MY3. The work of SM is supported by ANID FONDECYT Postdoctorado grant number 3240055.


\providecommand{\href}[2]{#2}\begingroup\raggedright\endgroup

\end{document}